\documentclass[ALICE,manyauthors]{cernphprep}
\newcommand{\pt}{p_\mathrm{T}} 

\newcommand{\ALICE}{\mbox{ALICE }}
\DeclareRobustCommand{\rchi}{{\mathpalette\irchi\relax}}
\newcommand{\irchi}[2]{\raisebox{\depth}{$#1\chi$}} 

\usepackage[comma,square,numbers,sort&compress]{natbib}
\usepackage{hyperref}
\usepackage{lineno}
\usepackage{xcolor}
\usepackage{xspace}
\usepackage{multirow}
\usepackage{makecell}
\usepackage{booktabs}
\usepackage{upgreek}
\usepackage[T1]{fontenc}

\begin{document}%

\begin{titlepage}
\PHyear{2021}
\PHnumber{057}      
\PHdate{12 April}  
%

\title{Energy dependence of $\phi$ meson production at forward rapidity\\
       in pp collisions at the LHC}
\ShortTitle{Energy dependence of $\phi$ meson production in pp collisions at the LHC}   

\Collaboration{ALICE Collaboration\thanks{See Appendix~\ref{app:collab} for the list of collaboration members}}
\ShortAuthor{ALICE Collaboration} 

\begin{abstract}
The production of $\phi$ mesons has been studied in pp collisions at LHC energies with the ALICE detector via the dimuon decay channel in the rapidity region $2.5 < y < 4$. 
Measurements of the differential cross section ${\rm d}^2\sigma/{\rm d}y {\rm d}\pt$ are presented as a function of the transverse momentum ($\pt$) at the center-of-mass energies $\sqrt{s}=5.02$, 8 and 13~TeV and compared with the ALICE results at midrapidity.
The differential cross sections at $\sqrt{s}=5.02$ and 13~TeV are also studied in several rapidity intervals as a function of $\pt$, and as a function of rapidity in three $\pt$ intervals. A hardening of the $\pt$-differential cross section with the collision energy is observed, while, for a given energy, $\pt$ spectra soften with increasing rapidity and, conversely, rapidity distributions get slightly narrower at increasing $\pt$. 
The new results, complementing the published measurements at $\sqrt{s}=2.76$ and 7~TeV, allow one to establish the energy dependence of $\phi$ meson production and to compare the measured cross sections with phenomenological models. None of the considered models manages to describe the evolution of the cross section with $\pt$ and rapidity at all the energies. 
\end{abstract}
\end{titlepage}
\setcounter{page}{2}

\section{Introduction}
\label{sect:Introduction}

Measurements of production cross sections and kinematic distributions of strange hadrons represent an effective tool to 
investigate strangeness production in high energy hadronic collisions, testing predictions from phenomenological models inspired by quantum chromodynamics (QCD).
In this context, the hardness of the specific partonic processes roughly separates two different regimes. 
At low transverse momentum ($\pt \lesssim 2$~GeV/$c$), non-perturbative processes dominate, described by phenomenological models where different approaches may be considered, such as the rope hadronization and color reconnection mechanisms implemented in PYTHIA~8~\cite{Skands:2014pea}. In this regime, strangeness production in hadronic collisions with no strange valence quark component in the initial state, such as proton--proton (pp), proton--nucleus and nucleus--nucleus collisions, depends on the s-quark content of the sea-parton wave function in nucleons. 
At high $\pt$ ($\pt \gtrsim 5$~GeV/$c$), strangeness production can typically be described in terms of hard 
partonic scattering processes via flavor creation and excitation, and gluon splitting, for which predictions can be obtained from perturbative calculations~\cite{Collins:1989gx}. The intermediate  $\pt$ region ($2<\pt<5$~GeV/$c$) is characterized by a smooth transition between the production mechanisms dominating at low
and high $\pt$, whose implementation in theoretical models is, however, not well defined and therefore data in this $\pt$ range are of particular importance. 

In addition, strangeness production is also addressed within the phenomenological statistical model approach~\cite{Andronic:2017pug}. 
In small hadronic systems, in particular, the effect of canonical suppression can play a significant role in determining the relative abundances of strange and lighter flavor hadrons, with the $\phi$ meson playing a special role due to its hidden strangeness composition~\cite{Kraus:2008fh, Kraus:2007hf}. 
A precise theoretical characterization of this mechanism in pp collisions, however, has yet to be established for the $\phi$ meson, and predictions are currently not available.

Measurements of $\phi$ meson production in small hadronic systems like pp collisions also provide the mandatory reference for the measurements in nucleus--nucleus collisions, where a precise pp baseline is needed to single out and characterize hot medium effects affecting particle production.

In this paper, results on the transverse momentum, rapidity and energy dependence of the $\phi$ meson production cross section at forward rapidity in pp collisions at the LHC energies are presented. 
Results are based on the data samples collected by ALICE (A~Large Ion Collider Experiment) at various energies during the LHC Run~1 and Run~2 and are compared with the predictions from the QCD-inspired models PYTHIA~8~\cite{Skands:2014pea}, PHOJET~\cite{Engel:1994vs, Engel:1995yda}, and EPOS~\cite{Werner:2012xh, Drescher200193, PhysRevC.89.064903}. 

The production of $\phi$ meson is studied by reconstructing the 2-body decay $\phi\to\mu^+\mu^-$ with the ALICE muon spectrometer. 
Results are reported in the forward rapidity ($y$) interval $2.5 < y < 4$ and for $\pt$ values ranging from 0.75 to 10~GeV/$c$, probing both soft and hard regimes of $\phi$ meson production.

\section{Experimental apparatus}
\label{sect:ExperimentalApparatus}

A full description of the \ALICE detector can be found in~\cite{Aamodt:2008zz, Abelev:2014ffa}. 
The results presented in this paper have been obtained using muon pairs detected with the forward muon spectrometer, which covers the pseudorapidity region $-4 < \eta< -2.5$. 
Throughout this paper, the sign of $\eta$ is determined by the choice of the ALICE reference system, while the kinematics of the reconstructed $\phi$ meson is referring to the positive rapidity hemisphere. 
The other detectors relevant for the current analysis are the Silicon Pixel Detector~(SPD) of the Inner Tracking System~(ITS), the V0 detector and the T0 detector.

The muon spectrometer is composed of a hadron absorber, followed by a set of tracking stations, a dipole magnet, an iron wall acting as muon filter, and a set of trigger stations. 
The hadron absorber, made of carbon, concrete and steel, is placed 0.9~m away from the interaction point. 
Its total material budget corresponds to 10 hadronic interaction lengths. 
The dipole magnet provides an integrated magnetic field of 3~$\mathrm{T}\,\mathrm{m}$ in the horizontal direction, perpendicular to the beam axis. 
The muon tracking is provided by five tracking stations, each one composed of two cathode pad chambers. 
The first two stations are located upstream of the dipole magnet, the third one in the middle of its gap and the last two downstream of it.
A 1.2~m thick iron wall, corresponding to 7.2~hadronic interaction lengths, 
is placed between the tracking and trigger detectors and  
absorbs the residual secondary hadrons emerging from the hadron absorber. 
The combined material budget of the hadron absorber and the iron wall stops muons with 
total momentum lower than $\sim 4$~GeV/$c$.
The muon trigger system consists of two stations, each one composed of two planes of resistive plate chambers (RPC), installed downstream of the muon filter. 
 
The SPD consists of two silicon pixel layers, covering the pseudorapidity regions $|\eta|<2.0$ and $|\eta|<1.4$ for the inner and outer layer, respectively. 
It is used for the determination of the primary interaction vertex position.
The V0 detector is composed of two scintillator hodoscopes covering the pseudorapidity regions $-3.7< \eta <-1.7$ and $2.8< \eta <5.1$.
The T0 detector is composed of two arrays of quartz Cherenkov counters, covering the pseudorapidity ranges $-3.3 < \eta < -3$ and $4.6 < \eta < 4.9$. 
The coincidence of a signal in both sides of either the T0 (8~TeV) or the V0 detectors (5.02 and 13~TeV) is used to define the minimum bias (MB) trigger and serves as input for the luminosity determination. It also allows for the offline rejection of beam--halo and beam--gas interactions.

\section{Data analysis}
\label{sect:DataAnalysis}

The analysis presented in this paper is based on the data samples collected by ALICE at $\sqrt{s} = 5.02,~8$ and 13~TeV. They complement the results already published at $\sqrt{s} = 2.76$~\cite{Adam:2015jca} and 7~TeV~\cite{ALICE:2011ad}.

\subsubsection*{Signal Extraction}

The data considered for the signal extraction were collected with a dimuon trigger, defined as the coincidence of a MB trigger and at least one pair of track segments reconstructed in the muon trigger system. The muon trigger system is configured to select muon tracks with a transverse momentum above a low-$\pt^\mu$ threshold, resulting in the conditions
$\pt^\mu\gtrsim 1$~GeV/$c$ for the data sample at $\sqrt{s} = 8$~TeV, and $\pt^\mu\gtrsim 0.5$~GeV/$c$ for the data samples at $\sqrt{s} = 5.02$ and 13~TeV\footnote{~Because of the design of the muon trigger system, the selection on the muon transverse momentum does not correspond to a sharp value. The reported values are the ones for which the trigger efficiency is $\sim 50$\,\%~\cite{Bossu:2012jt}. The different trigger thresholds are due to the fact that the data taking at $\sqrt{s}=8$~TeV was optimized for high invariant mass studies. The higher $\pt$ threshold was chosen to reduce the bandwidth in the data acquisition.}.
The number of dimuon trigger events thus selected are $\sim2 \times 10^{7}$ at $\sqrt{s} = 5.02$~TeV, $\sim8.4 \times 10^{5}$ at $\sqrt{s} = 8$~TeV and $\sim 2.8 \times 10^{8}$ at $\sqrt{s} = 13$~TeV.

Track reconstruction in the muon spectrometer is based on a Kalman filter algorithm~\cite{ALICE-INT-2009-044, ALICE-INT-2003-002}. Tracks reconstructed in the tracking chambers are requested to match a track segment reconstructed in the trigger chambers. In order to remove the tracks close to the acceptance borders, the muon pseudorapidity is required to be within the interval $-4 < \eta_{\mu} < -2.5$. 
Dimuons are formed by combining a pair of selected muon tracks, and their rapidity is explicitly imposed to be within the range $2.5 < y < 4$. 

The opposite-sign dimuon invariant mass spectrum contains a contribution of both uncorrelated and correlated pairs. The former mainly comes from decays of pions and kaons, which constitute the combinatorial background. This background is evaluated with an event mixing technique, in which a muon coming from an event is paired with a muon from a different event, so that the resulting muon pairs are uncorrelated by construction. This technique is described in detail in~\cite{ALICE:2011ad}. 

The processes contributing to the correlated dimuon mass spectrum in the low mass region, after combinatorial background subtraction, are the 2-body and Dalitz decays of the light resonances ($\eta$, $\rho$, $\omega$, $\eta'$, and $\phi$), usually referred to as the hadronic cocktail, superimposed onto a continuum mainly originating from semi-muonic decays of charm and beauty hadrons. 
In extracting the $\phi \to \mu\mu$ signal, however, no attempt is made to describe the underlying correlated continuum in terms of open charm and open beauty processes~\cite{ALICE-PUBLIC-2021-003}.  
Instead, a fit is applied to the mass spectrum after subtraction of the hadronic cocktail with an empirical function chosen among three options, all providing data description of similar quality: a polynomial of an appropriate degree, a single or double exponential\footnote{~The double exponential function has the form $f(m) = (m-2m_\mu) (A_1 e^{-m/m_1} + A_2 e^{-m/m_2})$, where $A_1$, $A_2$, $m_1$, $m_2$ are the free parameters of the fit.} and a Gaussian function whose width varies as a function of the mass as $\sigma(m) = \sigma_0 (1-e^{-\alpha m})$ and which will be referred to as ``variable-width Gaussian'' in the following. 
The degree of the polynomial is chosen to be the lowest one that satisfactorily fits the data, according to a statistical criterion based on the F-test, which can be briefly described as follows. 
The correlated continuum is first fitted with two polynomials of degree $n$ and $n+1$ respectively. The null hypothesis is that the continuum is equivalently described by the lower and the higher degree functions. 
This hypothesis is tested by applying the F-test to the results of the fit. If the resulting $p$-value is larger than 5\%, the lowest degree polynomial is used, otherwise the null hypothesis is rejected and the higher degree polynomial is chosen. 
The procedure starts from $n=3$ and is iterated until the F-test fails. The degree of the polynomial determined by means of the F-test is typically $n=4$. 
The F-test criterion is also applied, properly adapted, to choose among the single or double exponential options.

\begin{figure}[t!]
	\centering
	\includegraphics[width=0.49\textwidth]{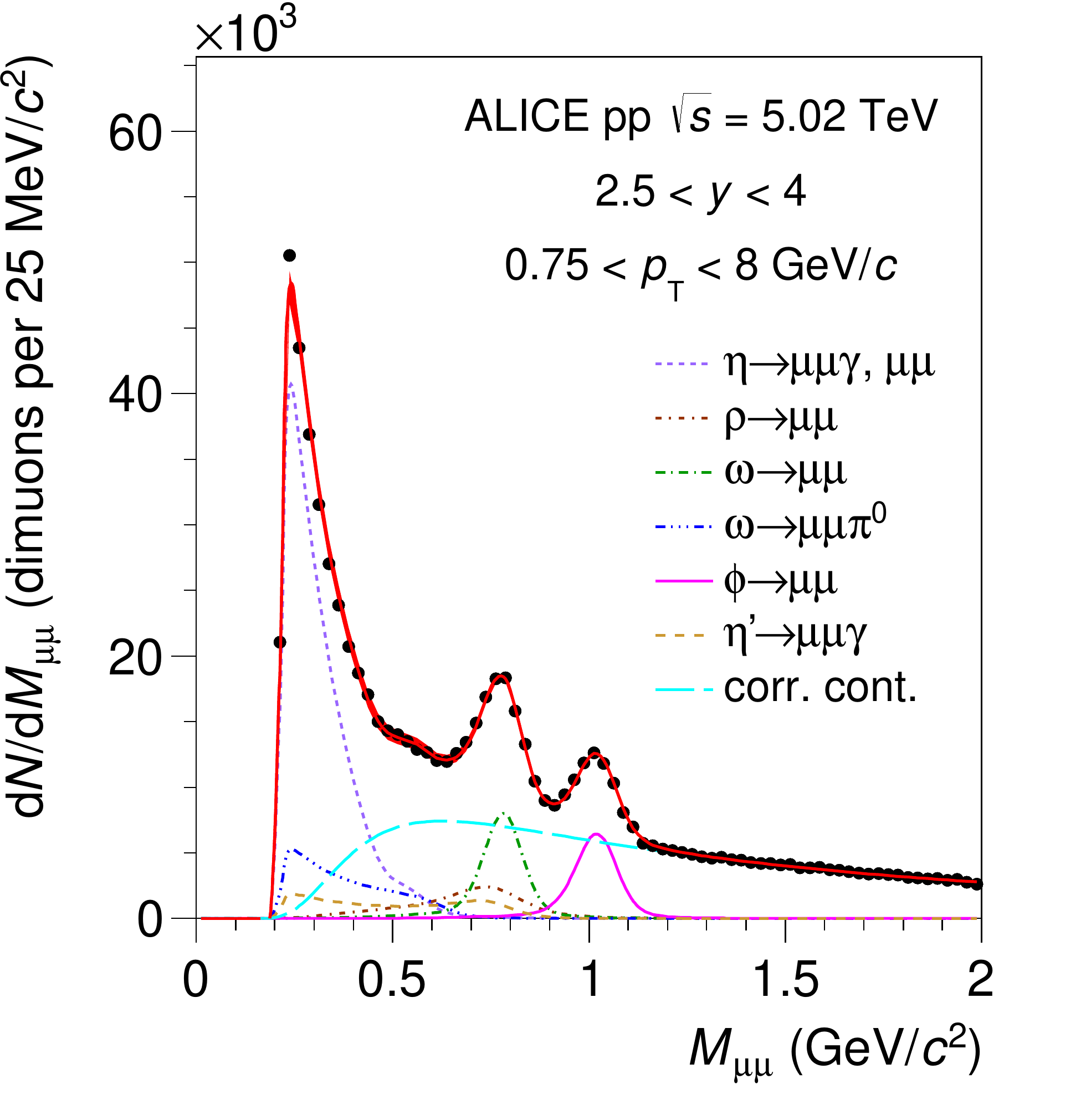}
    \includegraphics[width=0.49\textwidth]{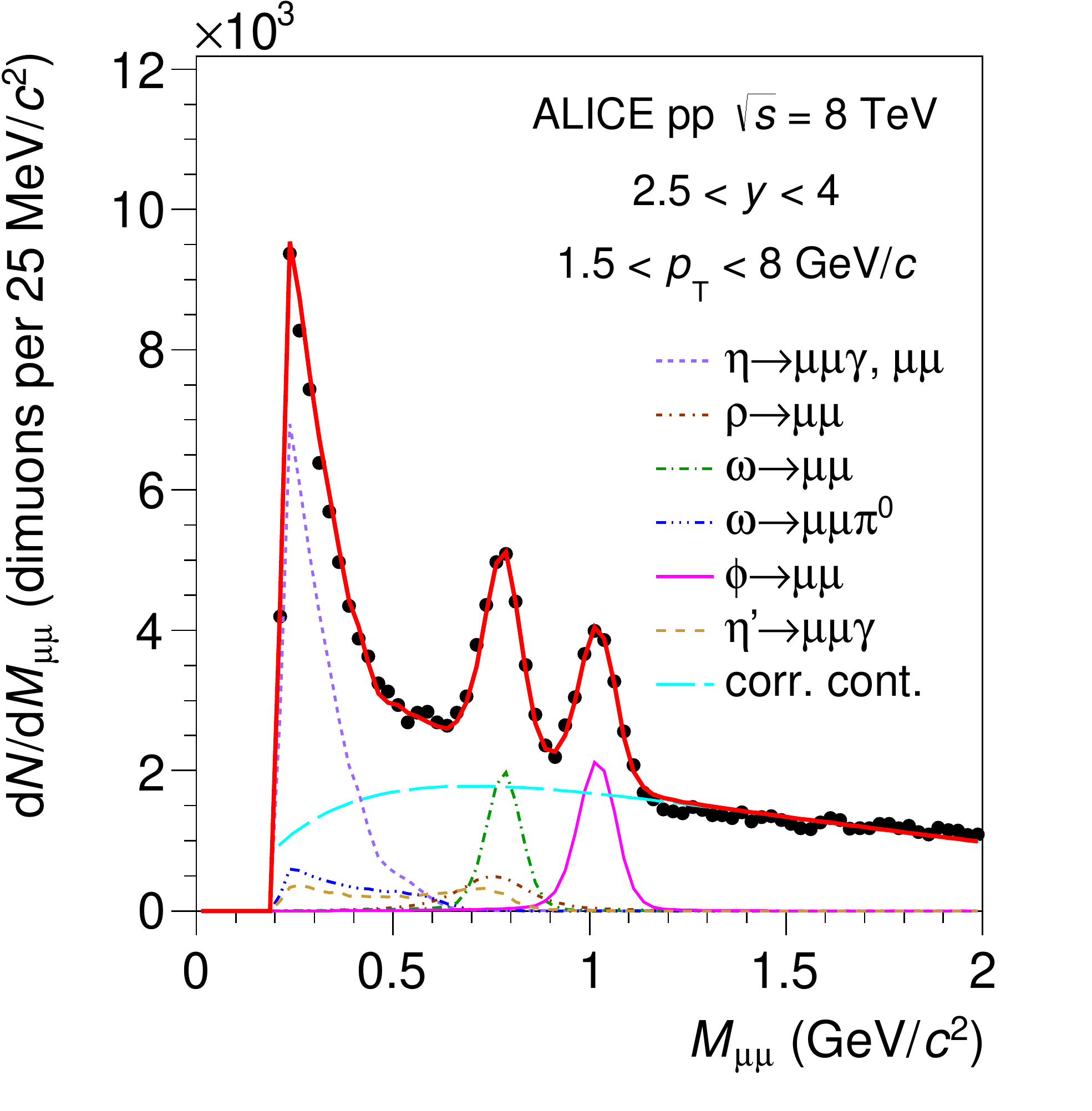}
    \includegraphics[width=0.49\textwidth]{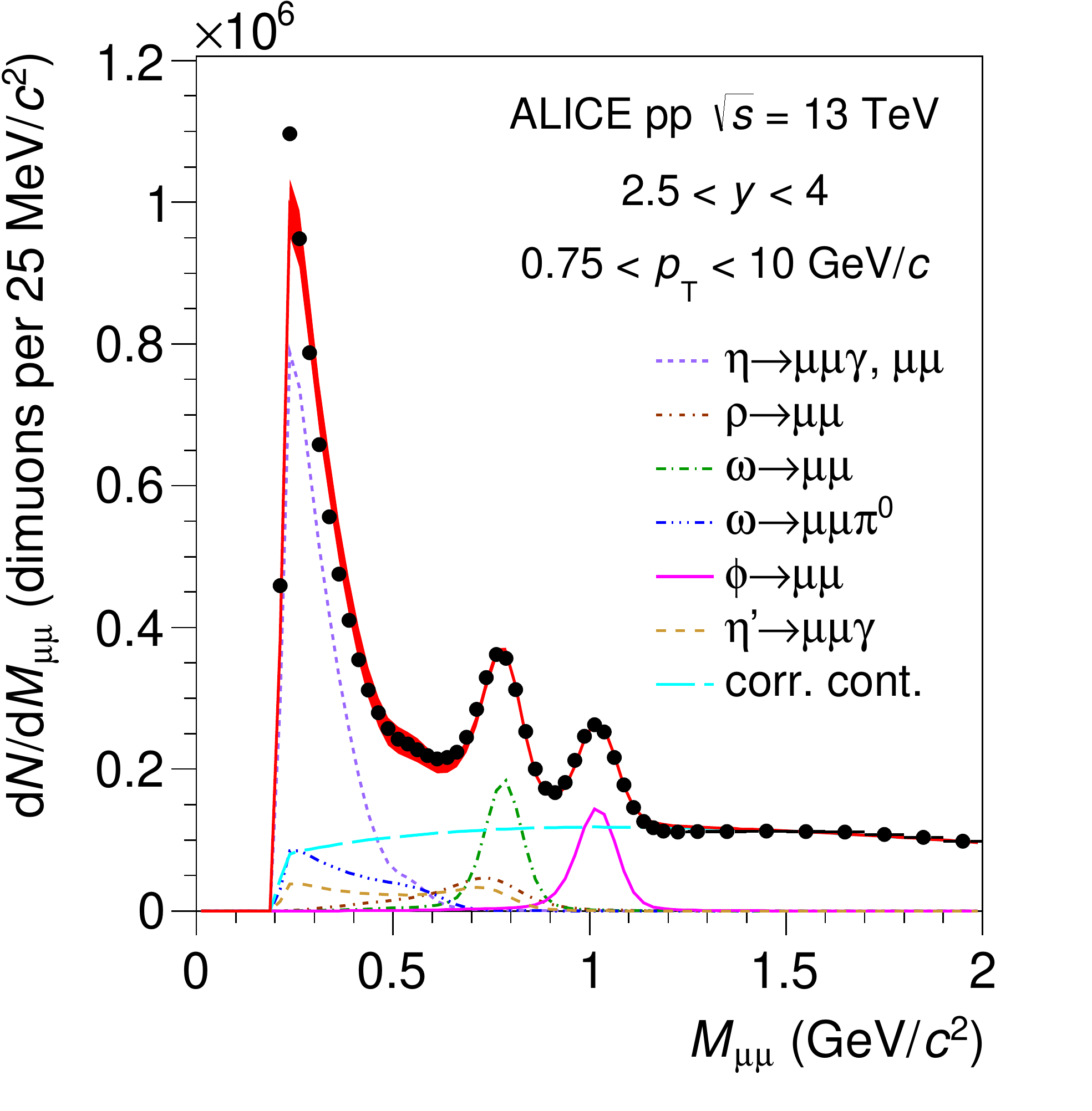}
	\caption{Examples of fits to the invariant mass spectra with the hadronic cocktail in pp collisions at $\sqrt{s} = 5.02$, 8 and 13~TeV, for muon pairs in $2.5 < y < 4$. The red histograms represent the sum of all simulated contributions. The correlated continuum has been fitted with a variable-width Gaussian function.}
	\label{fig:mass}
\end{figure}

The reconstructed opposite-sign dimuon mass spectrum is then fitted with a superposition of the hadronic cocktail and the regularized continuum discussed above: the procedure is applied independently for each of the three options considered for the empirical function describing the continuum. The free parameters of the fit are the normalization of the continuum and the $\eta \rightarrow \mu\mu \gamma$, $\omega \rightarrow \mu\mu$, and $\phi \rightarrow \mu\mu$ contributions, while the other processes are fixed according to the relative branching ratios or cross sections known from existing measurements~\cite{ALICE:2011ad,aguilar,refId0,Arnaldi:2019mgn}. 
The mass shapes of the processes included in the hadronic cocktail are extracted through full Monte Carlo (MC) simulations that include the resolution effects induced by the apparatus. The raw number of $\phi$ for each $(\Delta \pt, \Delta y)$ interval is then calculated as the mean of the results from the fits with the three different descriptions of the continuum. The typical reduced $\chi^2$ of these fits is around unity. 
In Fig.~\ref{fig:mass} the raw dimuon invariant mass spectrum after combinatorial background subtraction is shown for $\sqrt{s}=5.02$, 8 and 13~TeV, in their respective $\pt$ ranges. The components of the fit are also shown. Additional details on the signal extraction procedure are reported in Ref.~\cite{ALICE-PUBLIC-2021-003}.

The differential cross section for the $\phi$ meson is calculated as
\begin{equation*}
\label{eq:cross_section}
  \frac{\mathrm{d}^2\sigma_\phi}{\mathrm{d} y \mathrm{d}\pt} = \frac{1}{\Delta y \Delta \pt} \times \frac{N^{\mathrm{raw}}_{\phi\to\mu\mu}(\Delta\pt,\Delta y)}
    {[A\times\varepsilon](\Delta\pt,\Delta y)\times BR_{\phi\to\mu\mu}\times L_\mathrm{int}}~,
\end{equation*}
where $N^{\mathrm{raw}}_{\phi\to\mu\mu}(\Delta\pt,\Delta y)$ is the raw number of dimuons in the $\phi\to\mu\mu$ decay channel in a given
$\Delta \pt, \Delta y$ interval as obtained from the fit procedure described above, $[A\times\varepsilon](\Delta\pt,\Delta y)$ is the corresponding product of the geometrical acceptance and the reconstruction efficiency, $BR_{\phi\to\mu\mu}$ is the branching ratio for the $\phi\to\mu\mu$ decay, and $L_\mathrm{int}$ is the integrated luminosity of the analyzed data sample. 
The $[A\times\varepsilon](\Delta\pt,\Delta y)$ factor is evaluated by means of MC simulations, where the generation of the $\phi \to \mu \mu$ process is based on a parametric generator that takes as input $\pt$ and $y$ distributions iteratively tuned to the results of the present analyses. In detail, a first set of $\pt$ and $y$ distributions, corresponding to the results of the measurements at $\sqrt{s} = 7$~TeV, is taken as an input to the calculation. The resulting $[A\times\varepsilon](\Delta\pt,\Delta y)$ values are then used to correct the raw distributions obtained from the fits of the invariant mass spectra at the different energies. The corrected distributions are then used as input for another $[A\times\varepsilon](\Delta\pt,\Delta y)$ calculation, until convergence is reached.
For the branching ratio of the $\phi \to \mu \mu$ decay, the value measured for the dielectron channel $BR_{\phi \rightarrow ee} = (2.954 \pm 0.030) \times 10^{-4}$~\cite{PhysRevD.98.030001}  is used instead of the one of the dimuon channel, assuming lepton universality (i.e.~electroweak interaction coupling to all leptons with the same strength), because the latter is affected by a larger uncertainty.
The integrated luminosity is evaluated for each data set  as $L_\mathrm{int}=N_{\mu\mu} \times F_{\mathrm{norm}} / \sigma_{\mathrm{MB}}$, where $N_{\mu\mu}$ is the number of analyzed opposite-sign dimuon triggered events, $F_{\mathrm{norm}}$ is the inverse probability to obtain an opposite-sign dimuon trigger in a MB-triggered event, and $\sigma_{\mathrm{MB}}$ is the MB cross section measured using the van der Meer scan method~\cite{vanDerMeer}. 
The resulting values are $L_\mathrm{int}(5.02~\mathrm{TeV}) = 1.19 \pm 0.03$~pb$^{-1}$~\cite{ALICE-PUBLIC-2018-014}, $L_\mathrm{int}(8~\mathrm{TeV}) = 2.32 \pm 0.06$~pb$^{-1}$~\cite{ALICE-PUBLIC-2017-002}, and $L_\mathrm{int}(13~\mathrm{TeV}) = 7.35 \pm 0.40$~pb$^{-1}$~\cite{ALICE-PUBLIC-2016-002}, where the quoted uncertainties are the systematic ones, as the statistical uncertainties are negligible.

\subsubsection*{Systematic Uncertainties}

The systematic uncertainties on the measurement of the $\phi$ meson cross section include the following contributions: signal extraction, $A \times \varepsilon$ factor, branching ratio and integrated luminosity. 

Three sources of systematic uncertainty have been considered for the evaluation of $N^{\mathrm{raw}}_{\phi\to\mu\mu}$: the choice of the function used to describe the correlated continuum, the choice of the fit range, and the uncertainty on the relative branching ratios or cross sections used to adjust some of the processes contributing to the hadronic cocktail. 

For the description of the correlated continuum, the three empirical functions described above were considered, providing descriptions of the data of equivalent quality. 

The stability of the results under the choice of the fit range was studied by modifying the default range $2m_\mu < M_{\mu\mu} < 2$~GeV/$c^{2}$. Two alternative upper limits were considered: $ 2m_\mu < M_{\mu\mu} < 1.8$~GeV/$c^{2}$ and $ 2m_\mu < M_{\mu\mu} < 2.2$~GeV/$c^{2}$. 

The third contribution to the systematic uncertainty on the signal extraction was determined by varying the normalization of the $\rho \to \mu\mu$, $\eta^\prime \to \mu\mu\gamma$, $\eta \to \mu\mu$ and $\omega \to \mu\mu\pi^0$  processes relative to the main contributions to the hadronic cocktail ($\eta \to \mu\mu\gamma$, $\omega \to \mu\mu$ and $\phi \to\mu\mu$), by modifying the relative branching ratios or cross sections. 
To test the sensitivity of the fit results on the normalization of the first two processes, the $\sigma_{\rho}/\sigma_{\omega}$ ratio was varied by 10\,\%~\cite{Arnaldi:2019mgn}, while the $\sigma_{\eta^\prime}/\sigma_{\eta}$ ratio was varied by 50\,\%, according to the differences between $\sigma_{\eta^\prime}/\sigma_{\phi}$ obtained with calculations performed with PHOJET and with the PYTHIA 6 tunes D6T~\cite{Field:2008zz}, ATLAS-CSC~\cite{2004AcPPB.35.433B}, Perugia~0 and Perugia~11~\cite{Skands:2010ak}.  
To account for the variation of the other two processes, the relative branching ratios of the 2-body and Dalitz decay channels of $\eta$ and $\omega$ mesons were varied by one standard deviation, taking into account the values and the uncertainties reported in Ref.~\cite{PhysRevD.98.030001}.

\begin{table}[]
    \centering
        \caption{Sources of systematic uncertainties in the measurement of the $\phi$ yield at various energies for rapidity-integrated ($y$-int.) and double-differential (2-diff.) analysis.}
    \begin{tabular}{c c||c|c||c||c|c}
	\multicolumn{2}{c||}{$\sqrt{s}$} & \multicolumn{2}{c||}{5~TeV} & {8~TeV} & \multicolumn{2}{c}{13~TeV} \\
    \cline{3-7}
    & & $y$-int. & 2-diff. & $y$-int. & $y$-int. & 2-diff. \\
    	\hline
\makecell{$\pt, y$ dependent,\\ bin-to-bin uncorr.} &   \multicolumn{1}{|c||}{Extraction (\%)}	         & 3.7 & 3.1-4.2  & 2.1 & 3 - 12 & 2.6 - 20  \\
\hline
  \multirow{4}{*}{ \makecell{$\pt, y$ dependent,\\ bin-to-bin corr.}} & \multicolumn{1}{|c||}{Tracking (\%)}  & 2 & 2  & 7 & 4 & 4   \\
  & \multicolumn{1}{|c||}{Trigger (\%)} & 2.4-10.5 & 2.4-10.5 & 3.9-8.5 & 2.4 - 13 & 1 - 17   \\
   & \multicolumn{1}{|c||}{Matching (\%)} & 1 & 1 & 1 & 1 & 1   \\ 
    & \multicolumn{1}{|c||}{$A \times \varepsilon$ (\%)} & $<$1 & $<$1 & 3 & $<$1 & $<$1 \\ 
	\hline
     \multirow{2}{*}{ Global } &    \multicolumn{1}{|c||}{Branching Ratio (\%)}	         & 1 & 1 & 1 & 1 & 1  \\
   & \multicolumn{1}{|c||}{Luminosity (\%)}  & 2.3 & 2.3  & 2.6 & 5.1 & 5.1   \\    
    \hline
   \end{tabular}
    \label{tab:Syst_Contrib}
\end{table}

The total systematic uncertainty on $N^{\mathrm{raw}}_{\phi\to\mu\mu}$ is evaluated as the RMS of the values resulting from the above tests. At $\sqrt{s} = 5.02$ and 8~TeV the variations of this contribution as a function of $\pt$ are found to be small, so their mean value is assumed as final systematic uncertainty for the signal extraction. 
At $\sqrt{s} = 13$~TeV, the systematic uncertainty takes larger values in the lowest region of the measured $\pt$ range, mainly due to a progressive worsening in the description of the correlated continuum above the $\phi$ meson mass.

A potential source of systematic uncertainty on the correction for geometrical acceptance and reconstruction efficiency is associated to the choice of the input kinematic distributions used to generate the $\phi$ meson in the MC simulations. 
However, at $\sqrt{s}=5.02$ and $13$~TeV these distributions are tuned to the measured data and the corrections are performed in sufficiently small $\pt$ and rapidity intervals. 
The resulting uncertainty, evaluated by varying the parameters of the $\phi$ meson kinematic distributions, is found to be negligible. 
At $\sqrt{s}=8$~TeV only the $\pt$ distribution was tuned to the data. 
To evaluate the uncertainty due to the input rapidity distribution, alternative MC simulations at $\sqrt{s}=8$~TeV were performed using as input the rapidity distributions measured at $\sqrt{s}=5.02$ and $13$~TeV, and the corresponding $A\times\varepsilon$ factors were calculated. The half-difference between the two results is taken as the uncertainty due to the input rapidity distribution at $\sqrt{s}=8$ TeV and amounts to 3\%. 

In addition to that, three specific sources of systematic uncertainty were considered for the reconstruction efficiency: tracking efficiency, trigger efficiency and matching efficiency. 
The tracking efficiency was evaluated both from data and MC simulations. 
To this purpose, the MC was tuned to the detector condition during data taking, in order to reproduce the uncertainties arising from correlated and anticorrelated dead areas in the muon tracker. As the same tracking algorithm is applied on MC and data, the difference observed between the two estimates is assumed as the systematic uncertainty on the tracking efficiency.
The uncertainty on the trigger efficiency is mainly related to the imperfections in the description of two effects in the MC simulations: the interaction of the muons with the hadron absorber and the muon filter, and the occupancy of the trigger chambers.
The uncertainty corresponding to the first effect was estimated as the difference between $A \times \varepsilon$ obtained from simulations implementing GEANT3 or GEANT4 as alternative transport codes. 
The uncertainty on the occupancy of the trigger chambers was evaluated comparing $A \times \varepsilon$ resulting from simulations where the MC signal was simulated either as generated or by embedding it in the environment of a real event. 
The uncertainty on the matching efficiency is related to the choice of the $\chi^2$ cut used to define the matching between the tracks reconstructed in the tracking system and the track segments reconstructed in the trigger chambers, and amounts to 1\,\% for all data samples.

The remaining contributions to the systematic uncertainty are the ones due to the branching ratio of the $\phi\to ee$ decay channel ($\sim$1\,\%)~\cite{PhysRevD.98.030001} and to the integrated luminosity (2.1\,\%, 2.4\,\% and 5.0\,\% at $\sqrt{s}=5.02,~8$ and 13~TeV respectively). Two terms contribute to the uncertainty on the luminosity: the uncertainty on the visible cross section evaluated with the van der Meer scan technique and the difference between the luminosity measured with the T0 and the V0 detectors~\cite{ALICE-PUBLIC-2018-014,ALICE-PUBLIC-2017-002,ALICE-PUBLIC-2016-002}, while the uncertainty on the normalization factor  $F_{\mathrm{norm}}$, evaluated by calculating it with two different methods, amounts to 1\% for all data samples.

The systematic uncertainties listed above depend on both transverse momentum and rapidity, with the exception of the ones on the branching ratio and integrated luminosity. 
The sources of systematic uncertainty characterized by a dependence on transverse momentum and rapidity are classified into bin-to-bin uncorrelated (signal extraction) and bin-to-bin correlated (tracking and trigger efficiency, $A \times \varepsilon$ estimation, matching efficiency).

The contributions from the different sources of systematic uncertainties are reported in Table~\ref{tab:Syst_Contrib}, for both the rapidity-integrated and the double-differential analyses.

\section{Results}
\label{sect:Results}

\begin{figure}[thp!]
	\centering
	\includegraphics[width=0.48\textwidth]{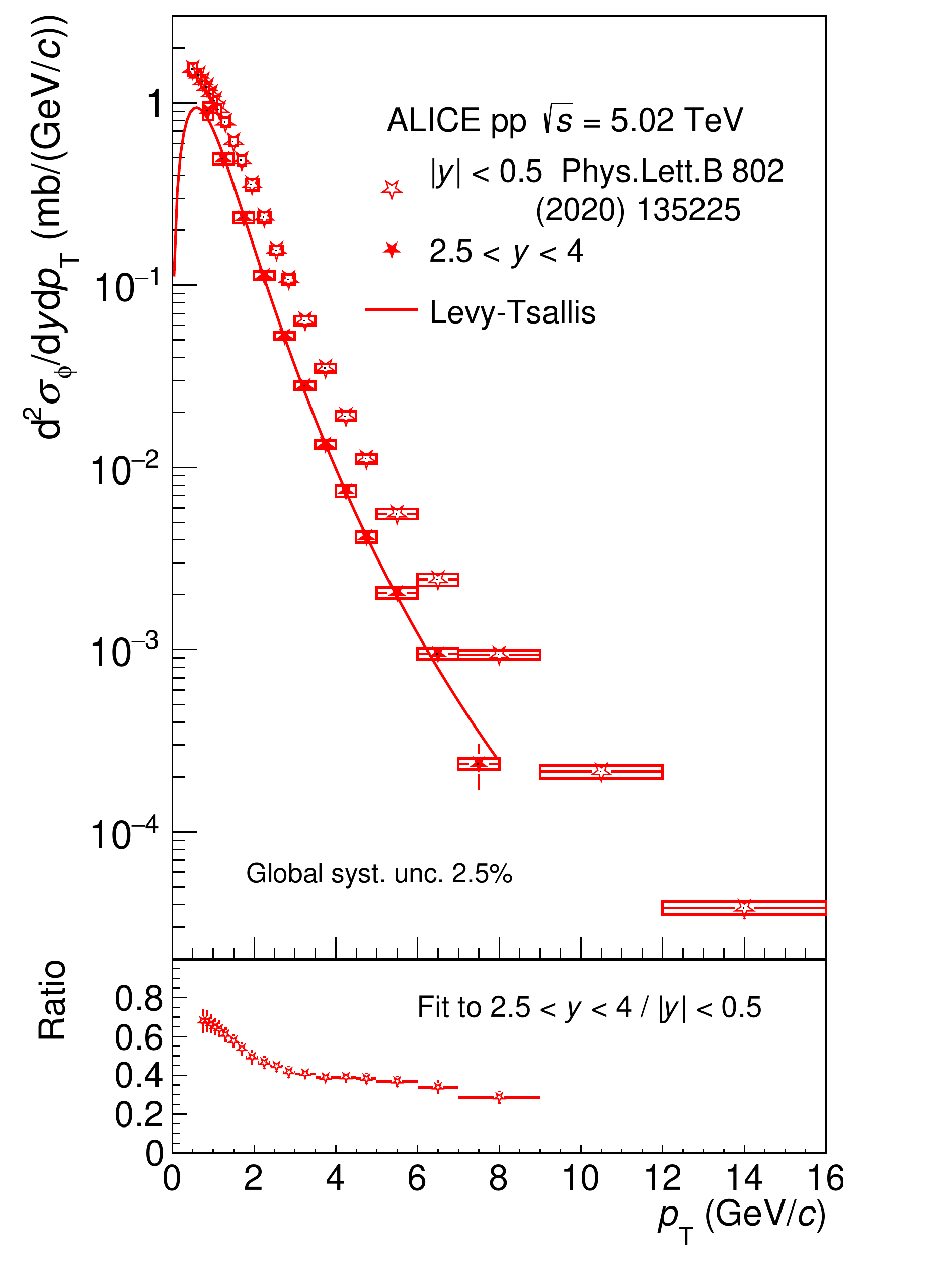}
	\includegraphics[width=0.48\textwidth]{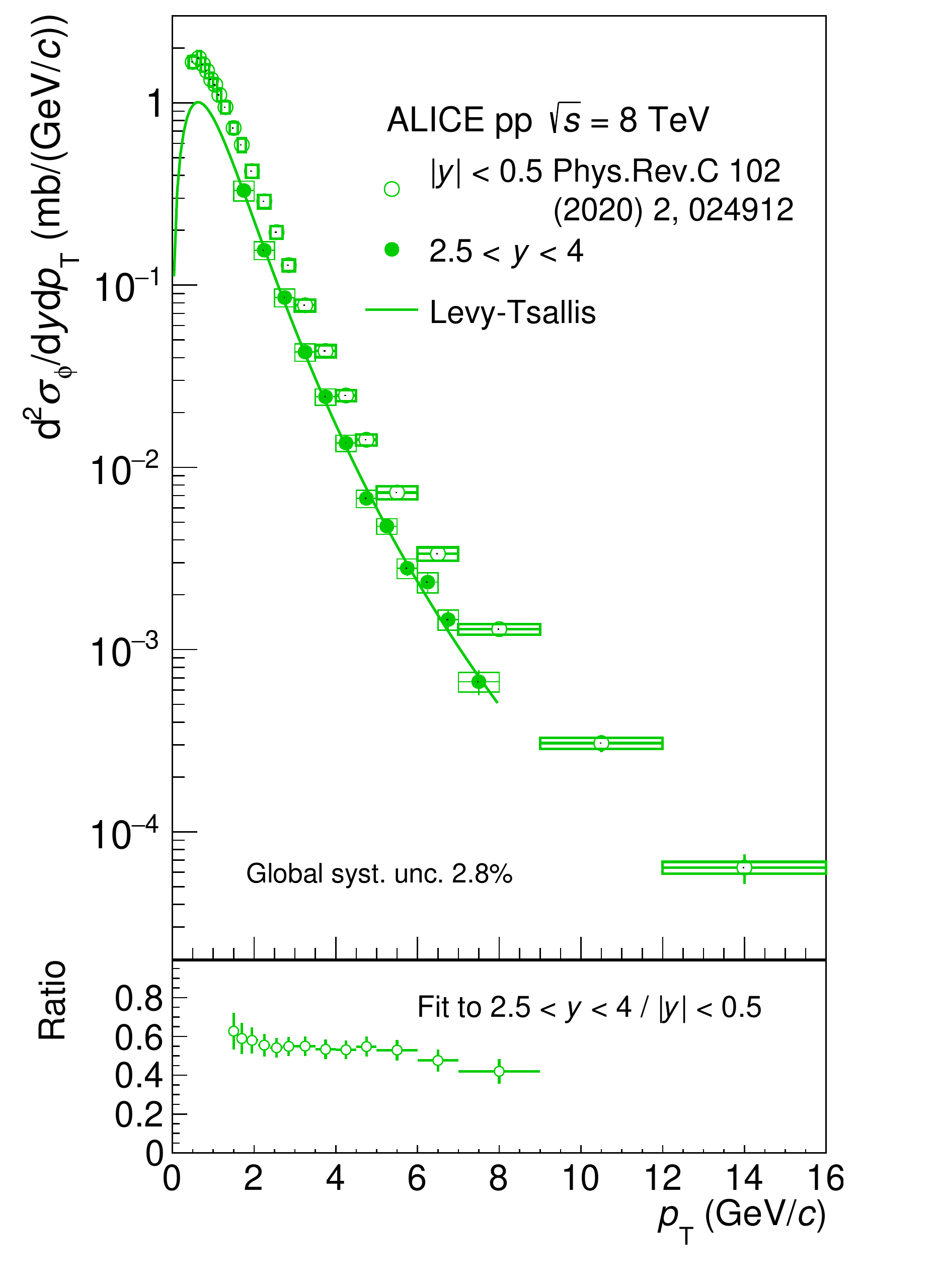}
    \includegraphics[width=0.48\textwidth]{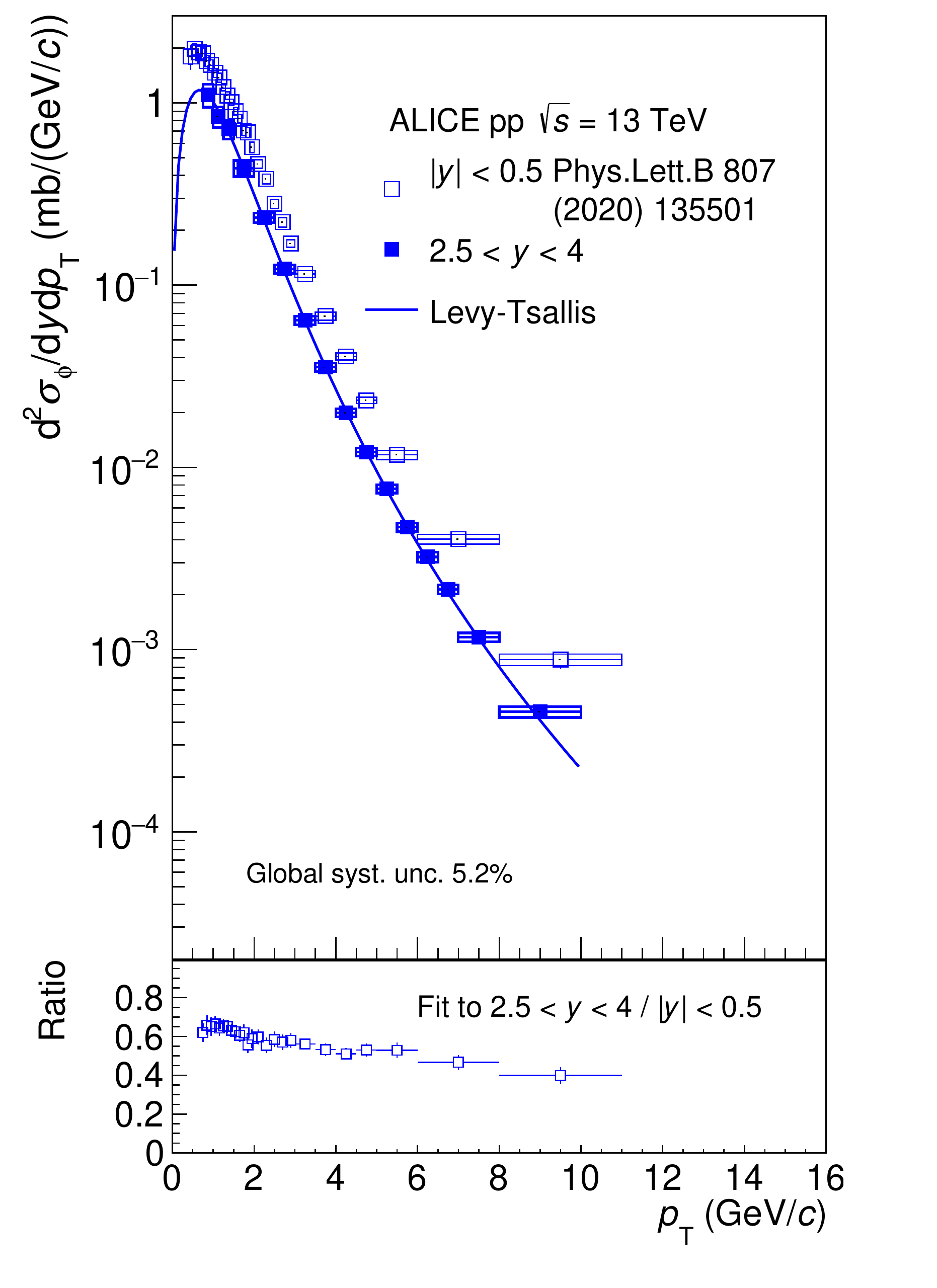}
	\caption{Differential $\phi$ meson production cross section $\mathrm{d}^2\sigma/(\mathrm{d}y\mathrm{d}\pt)$ as a function of $\pt$ at $\sqrt{s}=5.02$~TeV, $\sqrt{s}=8$~TeV and $\sqrt{s}=13$~TeV, measured in the $\mu^+\mu^-$ decay channel in the rapidity interval $2.5<y<4$ (full symbols) and in the K$^+$K$^-$ channel at midrapidity (open symbols). The boxes represent the systematic uncertainties, the error bars the statistical uncertainties. The data points are fitted with a Levy-Tsallis function. The ratio between the fit function at forward rapidity and the data at midrapidity is plotted in the bottom panels.}
	\label{fig:dsigmadydpt}
\end{figure}

\begin{figure}[tbh!]
	\centering
	\includegraphics[width=0.67\textwidth]{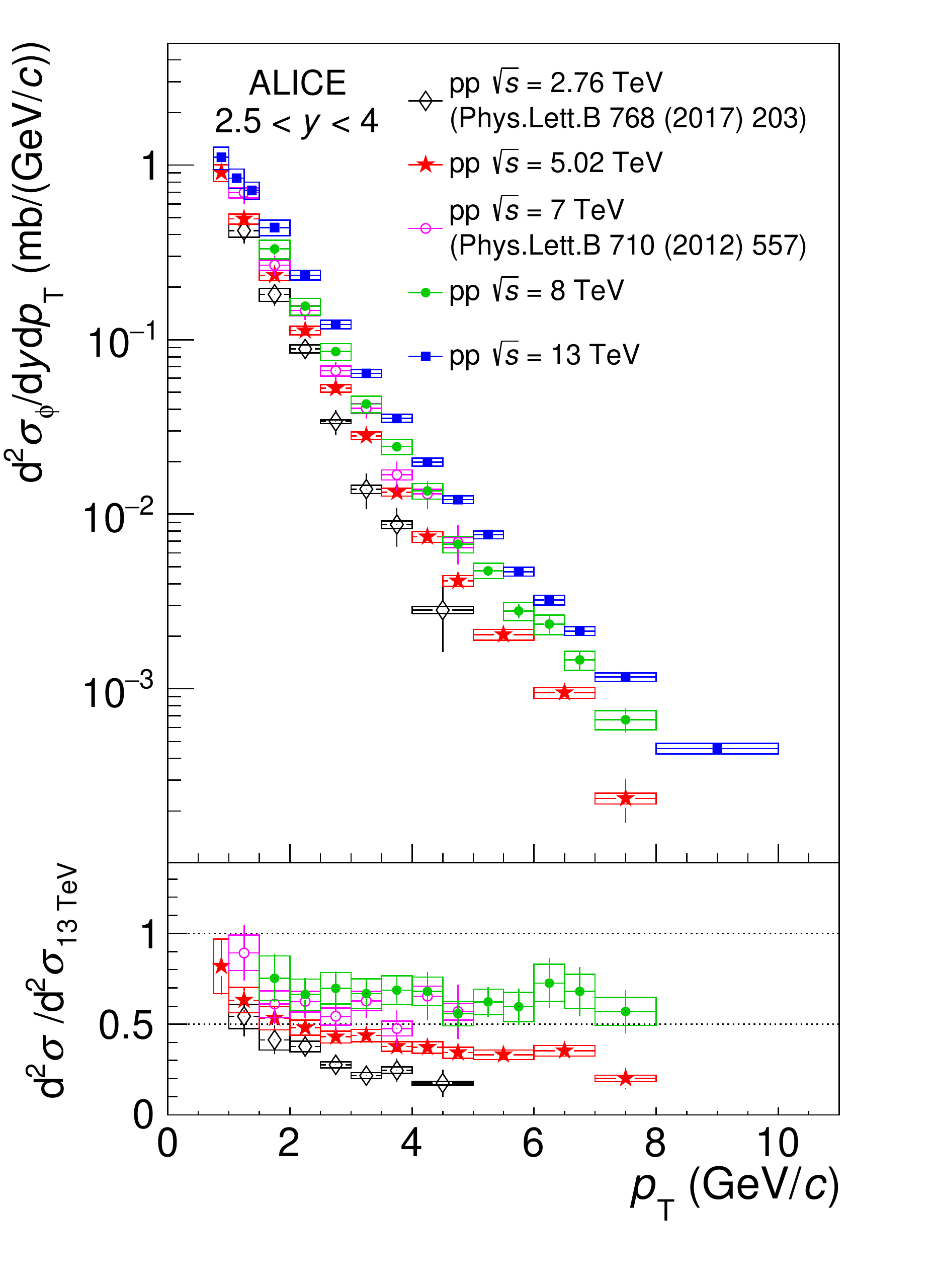}
	\caption{Top: differential $\phi$ meson production cross section $\mathrm{d}^2\sigma/(\mathrm{d}y\mathrm{d}\pt$) as a function of $\pt$ at $\sqrt{s}=$2.76, 5.02, 7, 8 and 13~TeV.
    The boxes represent the systematic uncertainties, the error bars the statistical uncertainties. 
    Bottom: ratio between the cross sections measured at several energies to the one obtained at $\sqrt{s}=13$~TeV.}
	\label{fig:ptspectra_all}
\end{figure}

The differential $\phi$ meson production cross section at $\sqrt{s}=5.02$, 8 and 13~TeV, measured in the rapidity range $2.5<y<4$, is shown in Fig.~\ref{fig:dsigmadydpt} as a function of the transverse momentum.
The cross sections are fitted with a Levy-Tsallis function~\cite{Tsallis:1987eu,Abelev:2006cs}:
\begin{equation*}
    f(\pt) = N_0 \frac{(n-1)(n-2)}{nT[nT+m(n-2)]}  \pt \left[1+ \frac{m_\mathrm{T} - m}{nT} \right]^{-n},
    \label{eq:LevyTsallis}
\end{equation*}
where $m_\mathrm{T}=  \sqrt{\pt^{2} + m_{\phi}^{2}}$ is the transverse mass and $N_0$, $n$ and $T$ are the free parameters of the fit. 
While the total systematic uncertainty is shown in the spectra of Fig.~\ref{fig:dsigmadydpt}, only the contribution coming from the signal extraction (added in quadrature to the statistical uncertainty) is considered when performing the fit, since the signal extraction is the only source of systematic uncertainty resulting in fully bin-to-bin uncorrelated fluctuations of the measured points. 

\begin{table}[]
    \centering
        \caption{Parameters of the Levy-Tsallis fits to the differential cross sections. The average $\pt$, calculated using the fit functions, is also reported.}
    \begin{tabular}{c|c|c|c|c}
    Rapidity interval       & $\rchi^{2}$/NdF &  $T$ (GeV)   & $n$            &   $\langle \pt \rangle$ (GeV/$c$) \\
    \hline
    \multicolumn{4}{c}{$\sqrt{s}=5.02$ TeV}  \\
    \hline
    $2.5 < y < 4$       & 1.41 & $0.273 \pm 0.011$ & $7.86 \pm 0.38$ &   $1.031 \pm 0.016$                  \\
    \hline
    $2.5 < y < 3$       & 1.57 & 0.273 (fixed) & $7.49 \pm 0.21$ &   $1.045\pm 0.009$                  \\
    $3 < y < 3.25$      & 0.95 & 0.273 (fixed) & $7.41 \pm 0.21$ &   $1.050 \pm 0.010$                  \\
    $3.25 < y < 3.5$    & 0.72 & 0.273 (fixed) & $7.85 \pm 0.22$ &   $1.029 \pm 0.009$                  \\
    $3.5 < y < 4$       & 1.91 & 0.273 (fixed) & $8.74 \pm 0.32$ &   $0.999 \pm 0.009$                  \\ 
    \hline
    \multicolumn{4}{c}{$\sqrt{s}=8$ TeV}  \\
    \hline
    $2.5 < y < 4$       & 0.52 & $0.310 \pm 0.045$ & $7.92 \pm 1.00$ &   $1.132 \pm 0.049$                  \\
    \hline
    \multicolumn{4}{c}{$\sqrt{s}=13$ TeV}  \\
    \hline
    $2.5 < y < 4$       & 1.13 & $0.341 \pm 0.005$ & $8.24 \pm 0.14$ &   $1.206 \pm 0.006$                  \\
    \hline
    $ 2.5 < y < 2.75$   & 2.03 & 0.341 (fixed) & $7.71 \pm 0.09$ &   $1.231 \pm 0.005$                  \\
    $ 2.75 < y < 3$     & 1.90 & 0.341 (fixed) & $8.45 \pm 0.09$ &   $1.196 \pm 0.004$                  \\
    $ 3 < y < 3.25$     & 0.92 & 0.341 (fixed) & $8.26 \pm 0.09$ &   $1.204 \pm 0.004$                  \\
    $ 3.25 < y < 3.5$   & 0.61 & 0.341 (fixed) & $8.54 \pm 0.11$ &   $1.192 \pm 0.004$                  \\
    $ 3.5 < y < 4$      & 0.75 & 0.341 (fixed) & $8.69 \pm 0.12$ &   $1.186 \pm 0.005$                  \\
    \hline
   \end{tabular}
    \label{tab:LevyTsallisFits}
\end{table}

The results of the fits are summarized in Table~\ref{tab:LevyTsallisFits}. The average $\pt$, calculated using the fit functions, increases by about 20\,\% when increasing the center-of-mass energy from 5.02 to 13~TeV.

In the same figure, the spectra at midrapidity measured by ALICE~\cite{Acharya:2019qge,Acharya:2019wyb,Acharya:2019bli}, normalized using the inelastic cross sections measured in~\cite{Loizides:2017ack}, are also reported for comparison. 
In the bottom panels, the ratio between the fits with the Levy-Tsallis functions at forward rapidity and the data at midrapidity is shown.
The cross section in the rapidity range covered by this analysis is approximately one half of the one measured at midrapidity. 
The $\pt$ spectra are harder at midrapidity. 
The difference between the slopes at forward and midrapidity is more evident at $\sqrt{s}=5.02$~TeV, the lowest energy considered. 

\begin{figure}[tbh!]
	\centering
	\includegraphics[width=0.999\textwidth]{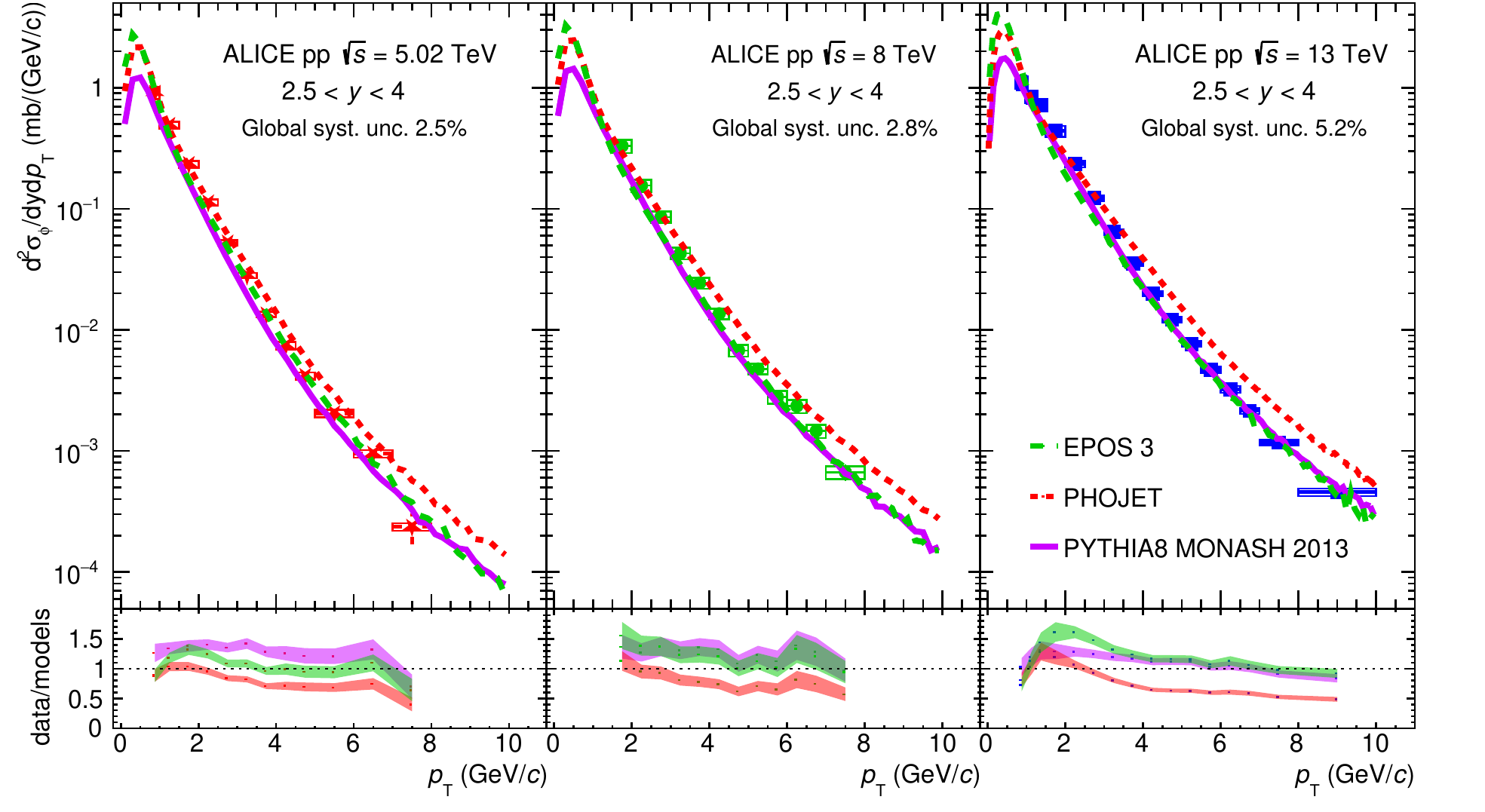}
	\caption{Top: differential $\phi$ meson production cross section $\mathrm{d}^2\sigma/\mathrm{d}y\mathrm{d}\pt$ as a function of $\pt$ at $\sqrt{s}=5.02$~TeV (left), $\sqrt{s}=8$~TeV (center) and $\sqrt{s}=13$~TeV (right) in the rapidity interval $2.5<y<4$, compared with EPOS~3~\cite{Werner:2012xh, Drescher200193, PhysRevC.89.064903}, PHOJET~\cite{Engel:1994vs, Engel:1995yda} and the Monash~2013 tune of PYTHIA~8.1~\cite{Skands:2014pea}.
	The boxes represent the systematic uncertainties, the error bars the statistical uncertainties.
	Bottom: ratio between the measured cross section and the calculations.}
	\label{fig:data_models}
\end{figure}

\begin{figure}[htb!]
	\centering
	\includegraphics[width=0.49\textwidth]{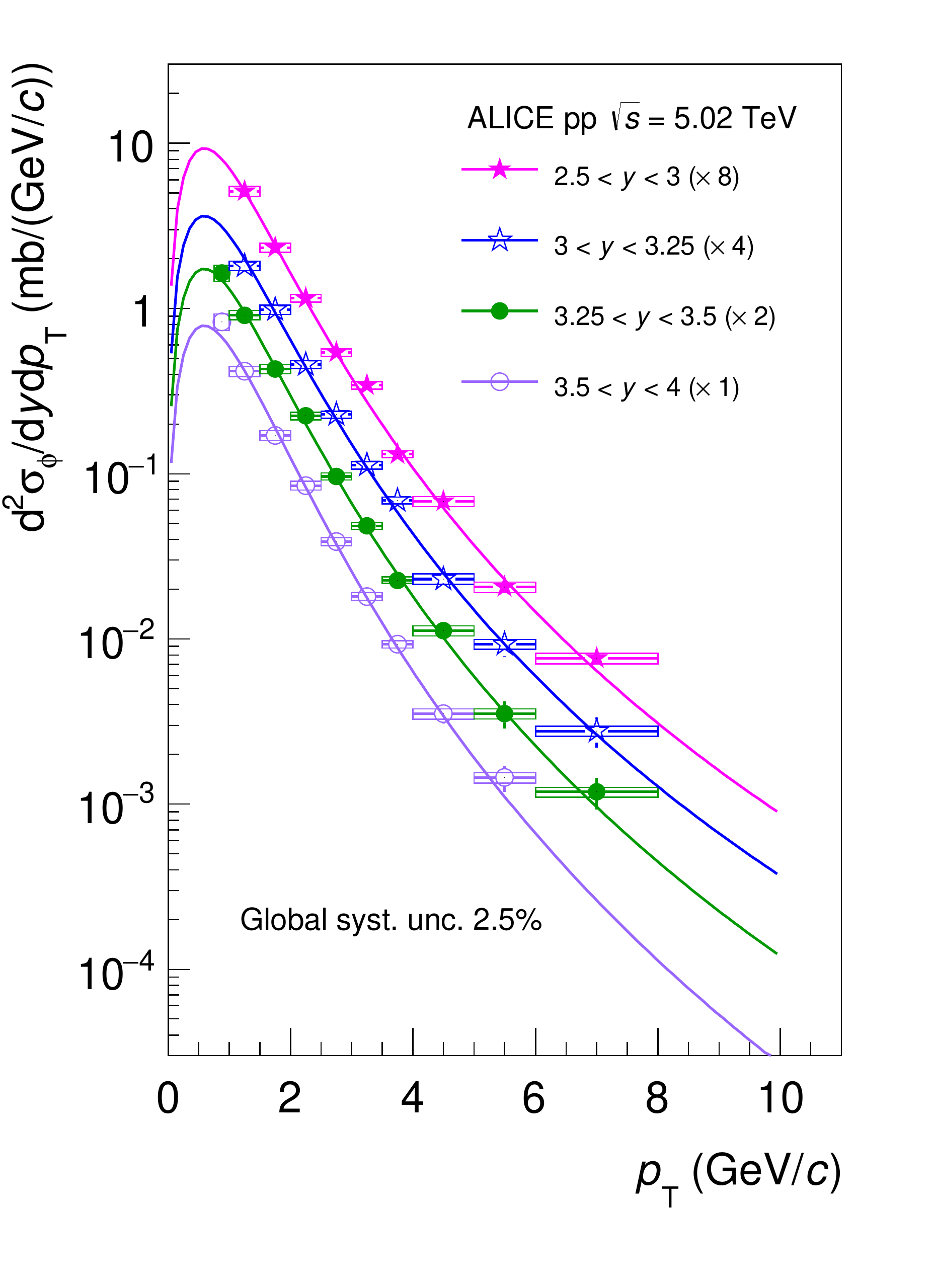}
    \includegraphics[width=0.49\textwidth]{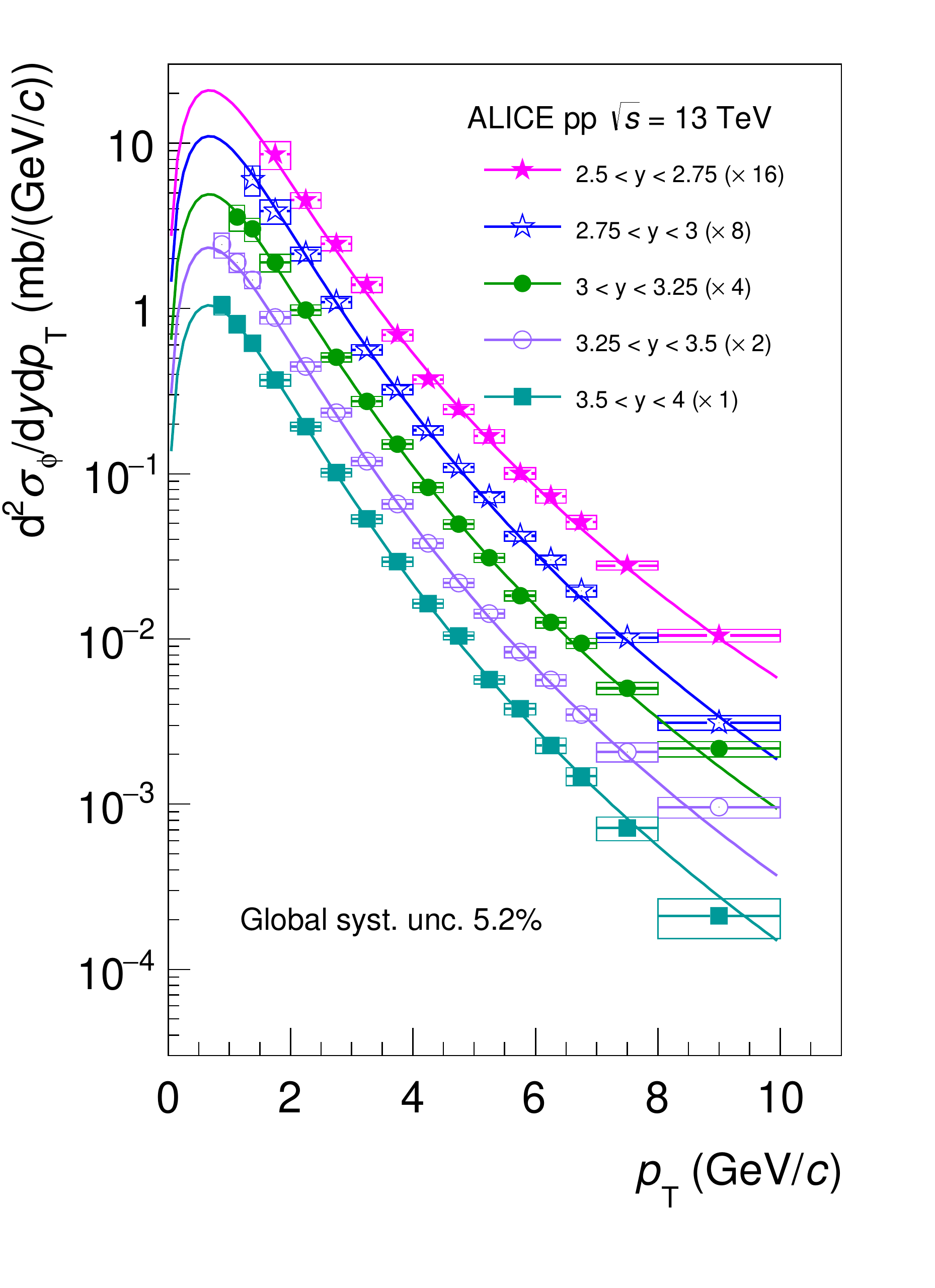}
    \vspace{-0.5cm}
	\caption{Differential $\phi$ meson production cross section $\mathrm{d}^2\sigma/\mathrm{d}y\mathrm{d}\pt$ as a function of $\pt$ at $\sqrt{s}=5.02$~TeV (left) and $\sqrt{s}=13$~TeV (right) in several rapidity intervals. 
	The boxes represent the systematic uncertainties, the error bars the statistical uncertainties.
	Data are fitted with a Levy-Tsallis function.}
	\label{fig:dsigmadydpt_ybins}
\end{figure}

In Fig.~\ref{fig:ptspectra_all}, the differential cross sections are compared with the previously published results at $\sqrt{s}=2.76$~TeV~\cite{Adam:2015jca} and $\sqrt{s}=7$~TeV~\cite{ALICE:2011ad}. 
The ratio to the measurement at $\sqrt{s}=13$~TeV is also reported in the bottom panel for a direct comparison. 
A hardening of the $\pt$ spectra is observed when increasing the center-of-mass energy. The values of the ratio at $\pt\sim 5$~GeV/$c$ change from $\sim 0.2$ for $\sqrt{s}=2.76$~TeV to $\sim0.65$ for $\sqrt{s}=8$~TeV.

In Fig.~\ref{fig:data_models} data are compared with the calculations performed with the models EPOS~3~\cite{Werner:2012xh, Drescher200193, PhysRevC.89.064903}, PHOJET~\cite{Engel:1994vs, Engel:1995yda} and the Monash~2013 tune of PYTHIA~8.1~\cite{Skands:2014pea}.
At all collision energies, EPOS~3 underestimates the cross section  at low $\pt$, while it describes the data for $\pt>4$~GeV/$c$. 
Vice versa, PHOJET reproduces the low-$\pt$ region up to $\pt\sim 2$~GeV/$c$, but does not describe the shape of the spectra, which is predicted to be harder by the model.
PYTHIA~8.1 with the Monash~2013 tune better reproduces the shape of the differential cross section at all energies. However, it still underestimates the measurement at $\sqrt{s}=5.02$ and 8~TeV, and reproduces the results at $\sqrt{s}=13$~TeV at high $\pt$ only, while it underestimates the data by about 20\% at low $\pt$.

\begin{figure}[bth!]
	\centering
	\includegraphics[width=0.49\textwidth]{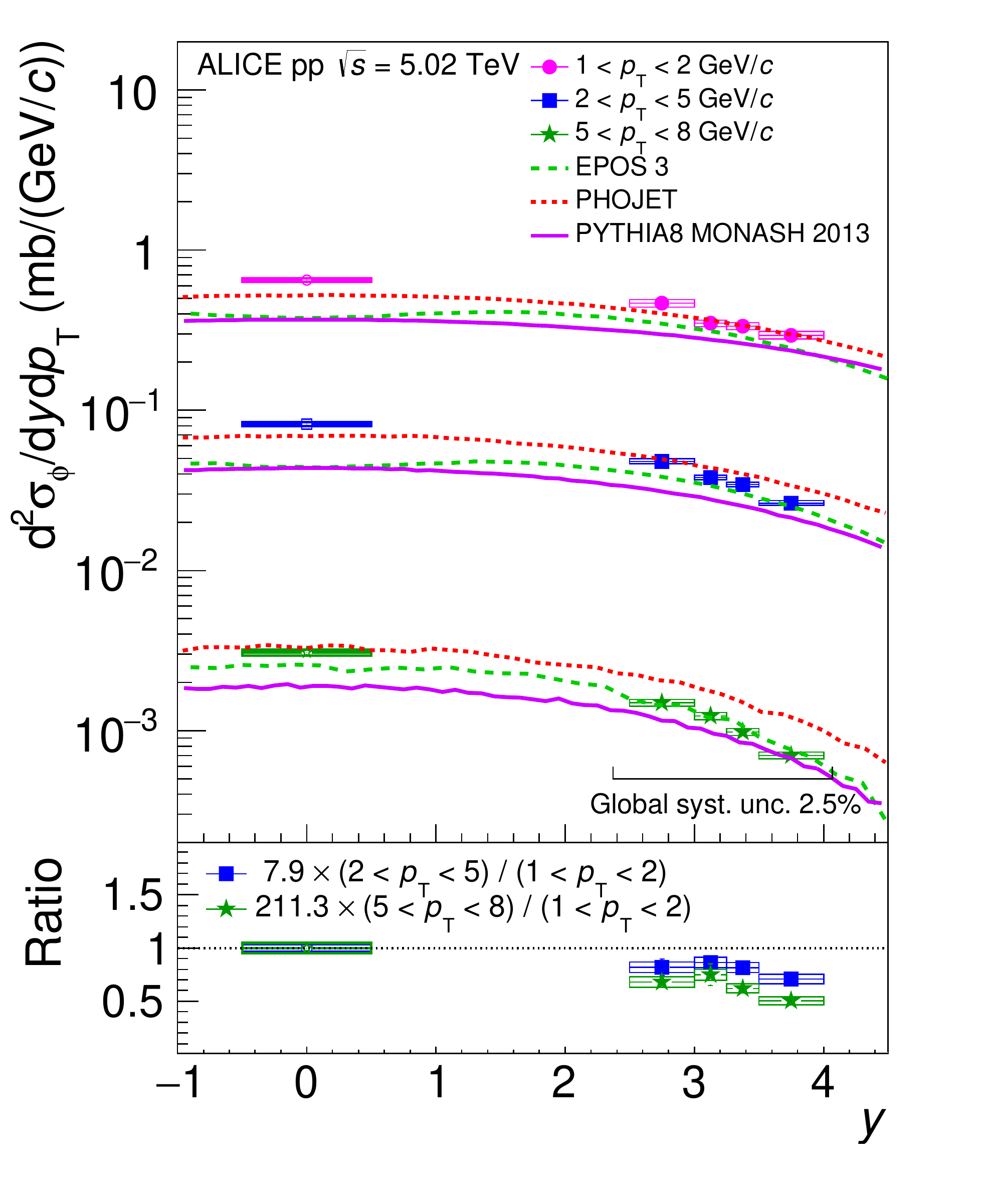}
    \includegraphics[width=0.49\textwidth]{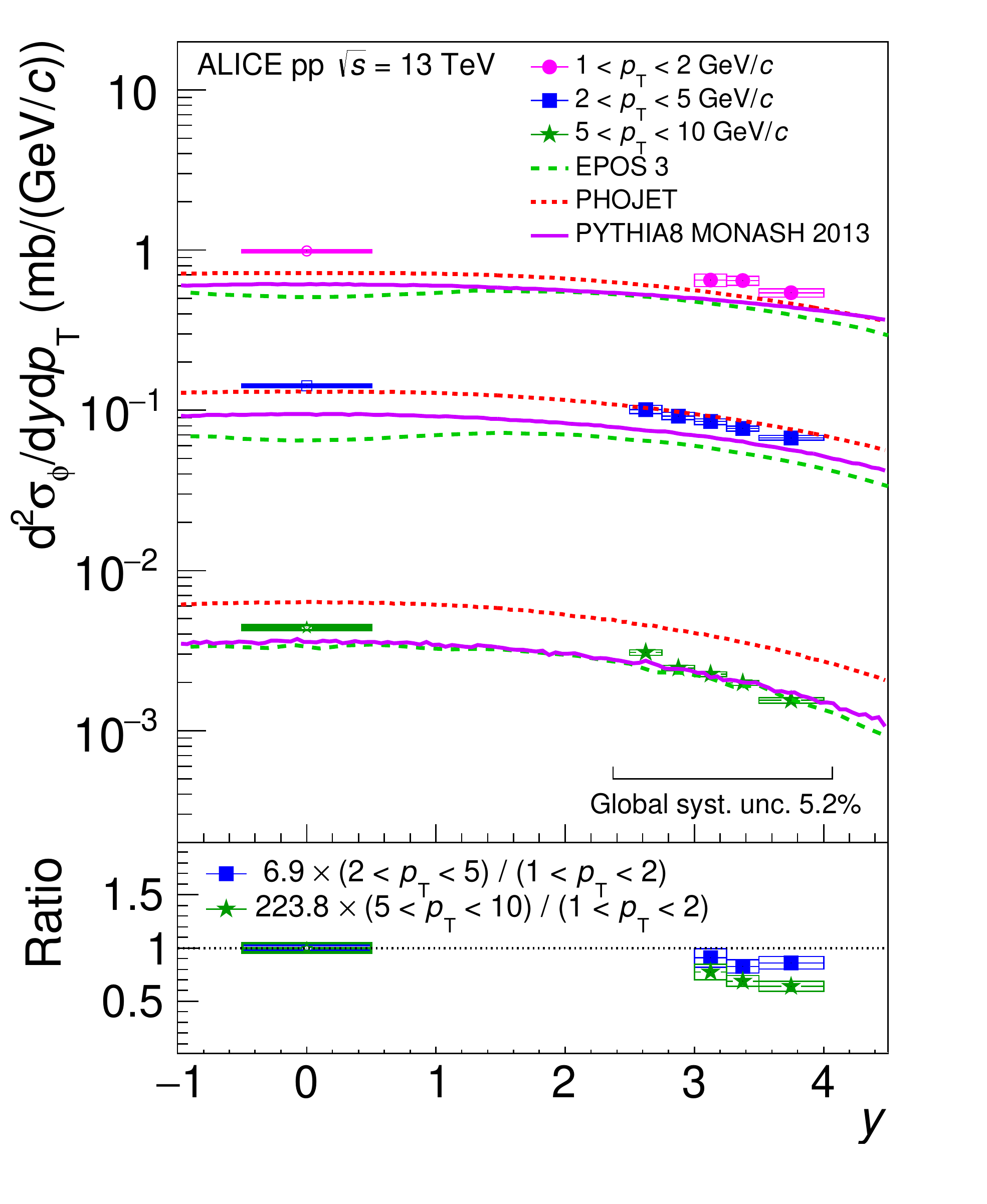}
    \vspace{-0.5cm}
	\caption{Differential $\phi$ meson production cross section $\mathrm{d}^2\sigma/\mathrm{d}y\mathrm{d}\pt$ as a function of rapidity at $\sqrt{s}=5.02$~TeV (left) and $\sqrt{s}=13$~TeV (right) in several $\pt$ intervals, compared with EPOS~3~\cite{Werner:2012xh, Drescher200193, PhysRevC.89.064903}, PHOJET~\cite{Engel:1994vs, Engel:1995yda} and the Monash~2013 tune of PYTHIA~8.1~\cite{Skands:2014pea}. The ratio of the data to the lowest $\pt$ interval is shown in the bottom panel.
    The boxes represent the systematic uncertainties, the error bars the statistical uncertainties.
	}
	\label{fig:dsigmadydpt_ptbins}
\end{figure}

At $\sqrt{s}=5.02$ and 13~TeV, the $\pt$ dependence of the differential cross section was also measured in several rapidity intervals. 
Results are shown in Fig.~\ref{fig:dsigmadydpt_ybins}. 
The $\pt$ coverage depends on the rapidity interval: in fact, at low $\pt$, $A \times \varepsilon$ significantly increases with rapidity; on the other side, at high $\pt$, high rapidity dimuons are more affected by statistical limitations than low rapidity ones. A significant dependence on $\pt$ and rapidity is also observed for the systematic uncertainty, namely for the contributions coming from the signal extraction and the trigger efficiency. Since this effect is related to the S/B and the data taking conditions, the  impact on the results depends on the considered data sample: this explains the slight difference in the $\pt$ and rapidity coverage of the measurements at $\sqrt{s} = 5$ and~13~TeV.
The $\pt-$differential cross sections are fitted with a Levy-Tsallis function, fixing the $T$ parameter to the value obtained from the fit in the full range $2.5<y<4$.
Fit results are reported in Tab.~\ref{tab:LevyTsallisFits}, together with the average $\pt$ calculated using the fit functions.
For both energies, $\sqrt{s}=5.02$ and 13~TeV, a moderate decrease of $\langle \pt \rangle$ as a function of rapidity is observed. However, the relatively large uncertainties do not allow to draw any firm conclusion on this trend.

To cross check the consistency between the results shown in Fig.~\ref{fig:dsigmadydpt} and~\ref{fig:dsigmadydpt_ybins}, the latter were integrated over the rapidity range $2.5<y<4$. 
The differences between the two methods in the common $\pt$ region amounts to about $5\,\%$. As a comparison, the systematic uncertainty on the cross section due to signal extraction is about $4\,\%$.

The differential cross sections measured at $\sqrt{s}=5.02$ and 13~TeV are shown in Fig.~\ref{fig:dsigmadydpt_ptbins} as a function of rapidity for several $\pt$ intervals, together with the corresponding values at midrapidity~\cite{Acharya:2019qge,Acharya:2019wyb,Acharya:2019bli}. 
The calculations performed with EPOS~3, PHOJET and PYTHIA~8.1 are also plotted. 
None of the considered models manages to reproduce the measured rapidity spectra in all the $\pt$ ranges, neither at 5 TeV nor at 13 TeV.

In the bottom panel of Fig.~\ref{fig:dsigmadydpt_ptbins}, the ratio of the rapidity distributions in a given $\pt$ interval to the lowest is shown, scaled such that the ratio at midrapidity is set to unity. 
A moderate narrowing of the rapidity distribution is observed when increasing the transverse momentum. This effect depends on the collision energy, being stronger for the lowest $\sqrt{s}$. 

\begin{figure}[t!]
	\centering
	\includegraphics[width=0.6\textwidth]{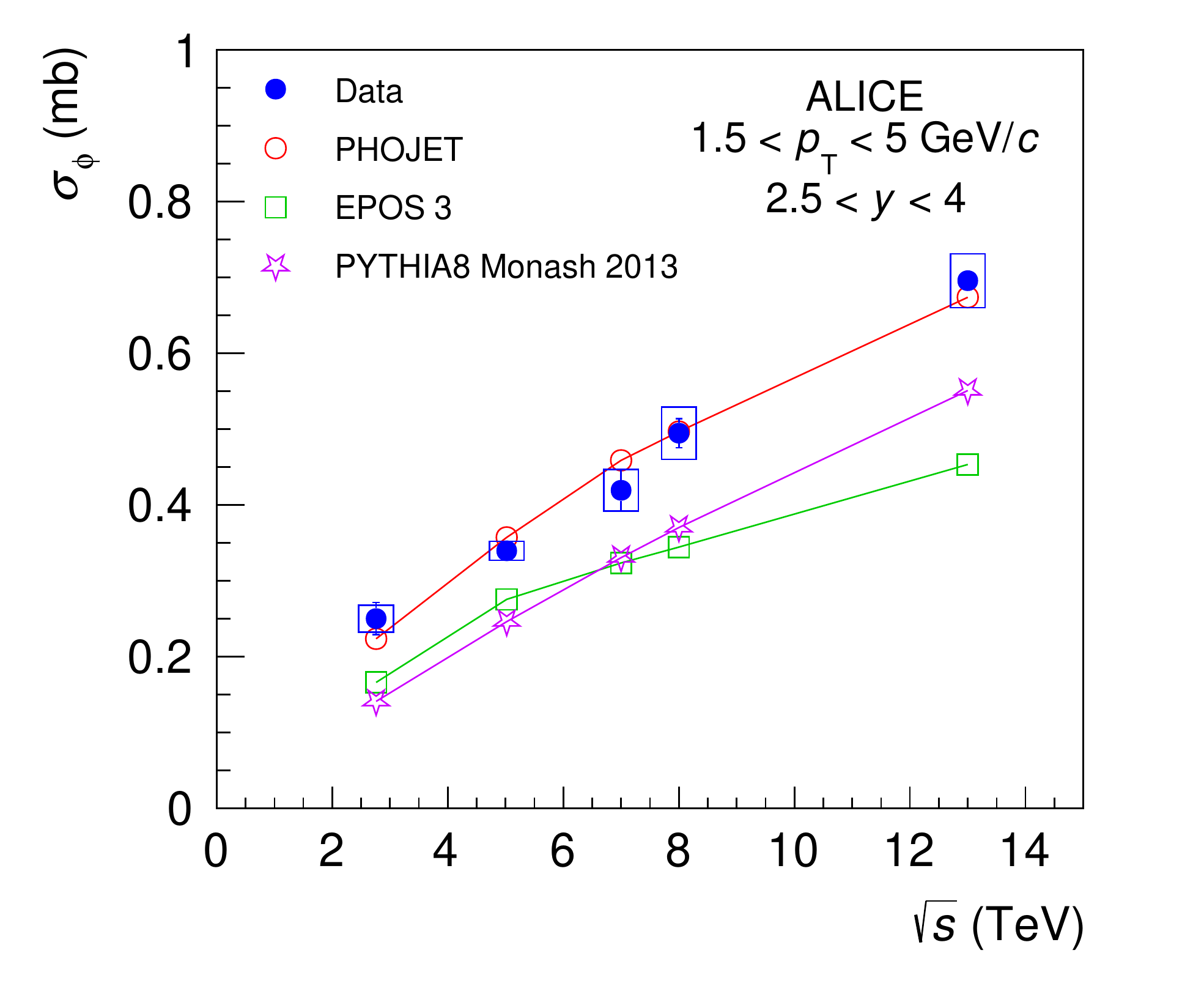}
	\caption{$\phi$ meson production cross section as a function of $\sqrt{s}$ for $1.5 < \pt <5$~GeV/$c$ and $2.5<y<4$, compared with EPOS~3~\cite{Werner:2012xh, Drescher200193, PhysRevC.89.064903}, PHOJET~\cite{Engel:1994vs, Engel:1995yda} and the Monash~2013 tune of PYTHIA~8.1~\cite{Skands:2014pea}. 
	The boxes represent the systematic uncertainties, the error bars the statistical uncertainties.
	}
	\label{fig:sigma_vs_sqrts}
\end{figure}

The $\phi$ meson production cross section integrated in the $\pt$ range $1.5<\pt<5$~GeV/$c$, common to the five energies, is plotted as a function of $\sqrt{s}$ in Fig.~\ref{fig:sigma_vs_sqrts}. 
Results are compared with EPOS~3, PHOJET and PYTHIA~8.1 with the Monash~2013 tune. 
In the $\pt$ interval considered for this study, the evolution of the cross section with the collision energy is well described by PHOJET. Both EPOS~3 and PYTHIA~8.1/Monash~2013 underestimate the absolute values by a factor ranging from about 1.2 to 1.7, while reproducing the trend as a function of the center-of-mass energy. The differences between the measurement and the calculations are mainly due to the overestimation of the cross section at the lowest $\pt$ accessible to the measurement.
\section{Conclusions}
\label{sect:Conclusions}

The $\phi$ meson production cross section was measured in pp collisions at the center-of-mass energies $\sqrt{s}=5.02,~8,~13$~TeV in the forward rapidity region $2.5<y<4$, complementing the previously published results at $\sqrt{s}=2.76$ and 7~TeV. 
The $\pt$ spectra are well described by a Levy-Tsallis function. 
A hardening of the $\pt$-differential cross section with the collision energy is observed, as is evinced from the comparison between the average values of the transverse  momentum or from the ratios between the differential cross sections as a function of $\pt$. At each energy, the $\pt$ spectra at midrapidity are harder than the corresponding ones at forward rapidity. 
Results were compared with the predictions from PYTHIA~8.1-Monash~2013, EPOS~3 and PHOJET.  
PHOJET reproduces the cross section at low $\pt$, while EPOS~3 better approaches the data for $\pt>4$~GeV/$c$. PYTHIA~8.1 with the Monash~2013 tune fairly describes the shape of the $\pt$-differential cross section at all energies and reproduces the results at $\sqrt{s}=13$~TeV, but underestimates the measurement at lower energies.

At $\sqrt{s}=5.02$ and 13~TeV, a double differential study of the $\phi$ meson production cross section was performed as a function of $\pt$ and rapidity.  None of the calculations manages to reproduce the rapidity spectra both at low and high $\pt$. 
A small decrease of the $\langle \pt \rangle$ value was observed with increasing rapidity, although with relatively large uncertainties. 
Analogously, the dependence of the cross section on rapidity appears to be slightly narrower when going towards higher $\pt$ values, thus showing that the correlation between $\pt$ and rapidity cannot be neglected in this rapidity range.

\newenvironment{acknowledgement}{\relax}{\relax}
\begin{acknowledgement}
\section*{Acknowledgements}

The ALICE Collaboration would like to thank all its engineers and technicians for their invaluable contributions to the construction of the experiment and the CERN accelerator teams for the outstanding performance of the LHC complex.
The ALICE Collaboration gratefully acknowledges the resources and support provided by all Grid centres and the Worldwide LHC Computing Grid (WLCG) collaboration.
The ALICE Collaboration acknowledges the following funding agencies for their support in building and running the ALICE detector:
A. I. Alikhanyan National Science Laboratory (Yerevan Physics Institute) Foundation (ANSL), State Committee of Science and World Federation of Scientists (WFS), Armenia;
Austrian Academy of Sciences, Austrian Science Fund (FWF): [M 2467-N36] and Nationalstiftung f\"{u}r Forschung, Technologie und Entwicklung, Austria;
Ministry of Communications and High Technologies, National Nuclear Research Center, Azerbaijan;
Conselho Nacional de Desenvolvimento Cient\'{\i}fico e Tecnol\'{o}gico (CNPq), Financiadora de Estudos e Projetos (Finep), Funda\c{c}\~{a}o de Amparo \`{a} Pesquisa do Estado de S\~{a}o Paulo (FAPESP) and Universidade Federal do Rio Grande do Sul (UFRGS), Brazil;
Ministry of Education of China (MOEC) , Ministry of Science \& Technology of China (MSTC) and National Natural Science Foundation of China (NSFC), China;
Ministry of Science and Education and Croatian Science Foundation, Croatia;
Centro de Aplicaciones Tecnol\'{o}gicas y Desarrollo Nuclear (CEADEN), Cubaenerg\'{\i}a, Cuba;
Ministry of Education, Youth and Sports of the Czech Republic, Czech Republic;
The Danish Council for Independent Research | Natural Sciences, the VILLUM FONDEN and Danish National Research Foundation (DNRF), Denmark;
Helsinki Institute of Physics (HIP), Finland;
Commissariat \`{a} l'Energie Atomique (CEA) and Institut National de Physique Nucl\'{e}aire et de Physique des Particules (IN2P3) and Centre National de la Recherche Scientifique (CNRS), France;
Bundesministerium f\"{u}r Bildung und Forschung (BMBF) and GSI Helmholtzzentrum f\"{u}r Schwerionenforschung GmbH, Germany;
General Secretariat for Research and Technology, Ministry of Education, Research and Religions, Greece;
National Research, Development and Innovation Office, Hungary;
Department of Atomic Energy Government of India (DAE), Department of Science and Technology, Government of India (DST), University Grants Commission, Government of India (UGC) and Council of Scientific and Industrial Research (CSIR), India;
Indonesian Institute of Science, Indonesia;
Istituto Nazionale di Fisica Nucleare (INFN), Italy;
Institute for Innovative Science and Technology , Nagasaki Institute of Applied Science (IIST), Japanese Ministry of Education, Culture, Sports, Science and Technology (MEXT) and Japan Society for the Promotion of Science (JSPS) KAKENHI, Japan;
Consejo Nacional de Ciencia (CONACYT) y Tecnolog\'{i}a, through Fondo de Cooperaci\'{o}n Internacional en Ciencia y Tecnolog\'{i}a (FONCICYT) and Direcci\'{o}n General de Asuntos del Personal Academico (DGAPA), Mexico;
Nederlandse Organisatie voor Wetenschappelijk Onderzoek (NWO), Netherlands;
The Research Council of Norway, Norway;
Commission on Science and Technology for Sustainable Development in the South (COMSATS), Pakistan;
Pontificia Universidad Cat\'{o}lica del Per\'{u}, Peru;
Ministry of Education and Science, National Science Centre and WUT ID-UB, Poland;
Korea Institute of Science and Technology Information and National Research Foundation of Korea (NRF), Republic of Korea;
Ministry of Education and Scientific Research, Institute of Atomic Physics and Ministry of Research and Innovation and Institute of Atomic Physics, Romania;
Joint Institute for Nuclear Research (JINR), Ministry of Education and Science of the Russian Federation, National Research Centre Kurchatov Institute, Russian Science Foundation and Russian Foundation for Basic Research, Russia;
Ministry of Education, Science, Research and Sport of the Slovak Republic, Slovakia;
National Research Foundation of South Africa, South Africa;
Swedish Research Council (VR) and Knut \& Alice Wallenberg Foundation (KAW), Sweden;
European Organization for Nuclear Research, Switzerland;
Suranaree University of Technology (SUT), National Science and Technology Development Agency (NSDTA) and Office of the Higher Education Commission under NRU project of Thailand, Thailand;
Turkish Energy, Nuclear and Mineral Research Agency (TENMAK), Turkey;
National Academy of  Sciences of Ukraine, Ukraine;
Science and Technology Facilities Council (STFC), United Kingdom;
National Science Foundation of the United States of America (NSF) and United States Department of Energy, Office of Nuclear Physics (DOE NP), United States of America.    
\end{acknowledgement}
\bibliographystyle{utphys}   

\bibliography{bibliography.bib}

\providecommand{\href}[2]{#2}\begingroup\raggedright\begin{thebibliography}{10}

\bibitem{Skands:2014pea}
P.~Skands, S.~Carrazza, and J.~Rojo, ``{Tuning PYTHIA 8.1: the Monash 2013
  Tune}'', \href{http://dx.doi.org/10.1140/epjc/s10052-014-3024-y}{{\em Eur.
  Phys. J.} {\bfseries C74} no.~8, (2014) 3024},
\href{http://arxiv.org/abs/1404.5630}{{\ttfamily arXiv:1404.5630 [hep-ph]}}.

\bibitem{Collins:1989gx}
J.~C. Collins, D.~E. Soper, and G.~F. Sterman, ``{Factorization of Hard
  Processes in QCD}'', \href{http://dx.doi.org/10.1142/9789814503266_0001}{{\em
  Adv. Ser. Direct. High Energy Phys.} {\bfseries 5} (1989) 1--91},
\href{http://arxiv.org/abs/hep-ph/0409313}{{\ttfamily arXiv:hep-ph/0409313
  [hep-ph]}}.

\bibitem{Andronic:2017pug}
A.~Andronic, P.~Braun-Munzinger, K.~Redlich, and J.~Stachel, ``{Decoding the
  phase structure of QCD via particle production at high energy}'',
  \href{http://dx.doi.org/10.1038/s41586-018-0491-6}{{\em Nature} {\bfseries
  561} no.~7723, (2018) 321--330},
\href{http://arxiv.org/abs/1710.09425}{{\ttfamily arXiv:1710.09425 [nucl-th]}}.

\bibitem{Kraus:2008fh}
I.~Kraus, J.~Cleymans, H.~Oeschler, and K.~Redlich, ``{Particle production in
  p-p collisions and prediction for LHC energy}'',
  \href{http://dx.doi.org/10.1103/PhysRevC.79.014901}{{\em Phys. Rev.}
  {\bfseries C79} (2009) 014901},
\href{http://arxiv.org/abs/0808.0611}{{\ttfamily arXiv:0808.0611 [hep-ph]}}.

\bibitem{Kraus:2007hf}
I.~Kraus, J.~Cleymans, H.~Oeschler, K.~Redlich, and S.~Wheaton, ``{Chemical
  Equilibrium in Collisions of Small Systems}'',
  \href{http://dx.doi.org/10.1103/PhysRevC.76.064903}{{\em Phys. Rev.}
  {\bfseries C76} (2007) 064903},
\href{http://arxiv.org/abs/0707.3879}{{\ttfamily arXiv:0707.3879 [hep-ph]}}.

\bibitem{Engel:1994vs}
R.~Engel, ``{Photoproduction within the two component dual parton model. 1.
  Amplitudes and cross-sections}'',
\href{http://dx.doi.org/10.1007/BF01496594}{{\em Z. Phys.} {\bfseries C66}
  (1995) 203--214}.

\bibitem{Engel:1995yda}
R.~Engel and J.~Ranft, ``{Hadronic photon-photon interactions at
  high-energies}'', \href{http://dx.doi.org/10.1103/PhysRevD.54.4244}{{\em
  Phys. Rev.} {\bfseries D54} (1996) 4244--4262},
\href{http://arxiv.org/abs/hep-ph/9509373}{{\ttfamily arXiv:hep-ph/9509373
  [hep-ph]}}.

\bibitem{Werner:2012xh}
K.~Werner, I.~Karpenko, M.~Bleicher, T.~Pierog, and S.~Porteboeuf-Houssais,
  ``{Jets, bulk matter, and their interaction in heavy ion collisions at
  several TeV}'', \href{http://dx.doi.org/10.1103/PhysRevC.85.064907}{{\em
  Phys. Rev. C} {\bfseries 85} (2012) 064907},
\href{http://arxiv.org/abs/1203.5704}{{\ttfamily arXiv:1203.5704 [nucl-th]}}.

\bibitem{Drescher200193}
H.~J. Drescher, M.~Hladik, S.~Ostapchenko, T.~Pierog, and K.~Werner, ``{Parton
  based Gribov-Regge theory}'',
  \href{http://dx.doi.org/10.1016/S0370-1573(00)00122-8}{{\em Phys. Rept.}
  {\bfseries 350} (2001) 93--289},
\href{http://arxiv.org/abs/hep-ph/0007198}{{\ttfamily arXiv:hep-ph/0007198
  [hep-ph]}}.

\bibitem{PhysRevC.89.064903}
K.~Werner, B.~Guiot, I.~Karpenko, and T.~Pierog, ``{Analyzing radial flow
  features in p-Pb and p-p collisions at several TeV by studying
  identified-particle production with the event generator EPOS3}'',
  \href{http://dx.doi.org/10.1103/PhysRevC.89.064903}{{\em Phys. Rev. C}
  {\bfseries 89} (2014) 064903},
\href{http://arxiv.org/abs/1312.1233}{{\ttfamily arXiv:1312.1233 [nucl-th]}}.

\bibitem{Aamodt:2008zz}
{\bfseries ALICE} Collaboration, K.~Aamodt {\em et~al.}, ``{The ALICE
  experiment at the CERN LHC}'',
\href{http://dx.doi.org/10.1088/1748-0221/3/08/S08002}{{\em JINST} {\bfseries
  3} (2008) S08002}.

\bibitem{Abelev:2014ffa}
{\bfseries ALICE} Collaboration, B.~Abelev {\em et~al.}, ``{Performance of the
  ALICE experiment at the CERN LHC}'',
  \href{http://dx.doi.org/10.1142/S0217751X14300440}{{\em Int. J. Mod. Phys. A}
  {\bfseries 29} (2014) 1430044},
\href{http://arxiv.org/abs/1402.4476}{{\ttfamily arXiv:1402.4476 [nucl-ex]}}.

\bibitem{Adam:2015jca}
{\bfseries ALICE} Collaboration, J.~Adam {\em et~al.}, ``{$\phi$-meson
  production at forward rapidity in p-Pb collisions at $\sqrt{s_{\rm NN}}$ =
  5.02 TeV and in pp collisions at $\sqrt{s}$ = 2.76 TeV}'',
  \href{http://dx.doi.org/10.1016/j.physletb.2017.01.074}{{\em Phys. Lett.}
  {\bfseries B768} (2017) 203--217},
\href{http://arxiv.org/abs/1506.09206}{{\ttfamily arXiv:1506.09206 [nucl-ex]}}.

\bibitem{ALICE:2011ad}
{\bfseries ALICE} Collaboration, B.~Abelev {\em et~al.}, ``{Light vector meson
  production in pp collisions at $\sqrt{s} = 7$~TeV}'',
  \href{http://dx.doi.org/10.1016/j.physletb.2012.03.038}{{\em Phys. Lett. B}
  {\bfseries 710} (2012) 557--568},
\href{http://arxiv.org/abs/1112.2222}{{\ttfamily arXiv:1112.2222 [nucl-ex]}}.

\bibitem{Bossu:2012jt}
{\bfseries ALICE} Collaboration, F.~Bossu, M.~Gagliardi, and M.~Marchisone,
  ``{Performance of the RPC-based ALICE muon trigger system at the LHC}'',
  \href{http://dx.doi.org/10.1088/1748-0221/7/12/T12002}{{\em JINST} {\bfseries
  7} (2012) T12002}, \href{http://arxiv.org/abs/1211.1948}{{\ttfamily
  arXiv:1211.1948 [physics.ins-det]}}.

\bibitem{ALICE-INT-2009-044}
{\bfseries ALICE} Collaboration, ``{Numerical Simulations and Offline
  Reconstruction of the Muon Spectrometer of ALICE}'', {\em ALICE-INT-2009-044}
  (2009) . \url{https://edms.cern.ch/document/1054937/1}.

\bibitem{ALICE-INT-2003-002}
{\bfseries ALICE} Collaboration, ``{Development of the Kalman filter for
  tracking in the forward muon spectrometer of ALICE}'', {\em
  ALICE-INT-2003-002} (2003) . \url{https://edms.cern.ch/document/371480/1}.

\bibitem{ALICE-PUBLIC-2021-003}
{\bfseries ALICE} Collaboration, ``{Signal extraction in the analysis of
  $\omega$ and $\phi$ meson production at forward rapidity in pp collisions at
  the LHC}'', {\em ALICE-PUBLIC-2021-003} (Apr, 2021) .
  \url{https://cds.cern.ch/record/2765657}.

\bibitem{aguilar}
{Aguilar-Benitez, M. et al.}, ``{Inclusive particle production in 400 GeV/c
  pp-interactions}'',
\href{http://dx.doi.org/10.1007/BF01551452}{{\em Z. Phys. C} {\bfseries 50}
  (1991) 405--426}.

\bibitem{refId0}
{G. Agakichiev et al.}, ``{Neutral meson production in p-Be and p-Au collisions
  at 450~GeV beam energy}'',
\href{http://dx.doi.org/10.1007/s100529800804}{{\em Eur. Phys. J. C} {\bfseries
  4} (1998) 249--257}.

\bibitem{Arnaldi:2019mgn}
{\bfseries NA60} Collaboration, R.~Arnaldi {\em et~al.}, ``{Nuclear dependence
  of light neutral meson production in p-A collisions at 400 GeV with NA60}'',
\href{http://dx.doi.org/10.1140/epjc/s10052-019-6848-7}{{\em Eur. Phys. J.}
  {\bfseries C79} no.~5, (2019) 443}.

\bibitem{PhysRevD.98.030001}
{\bfseries Particle Data Group} Collaboration, M.~e.~a. Tanabashi, ``Review of
  particle physics'', \href{http://dx.doi.org/10.1103/PhysRevD.98.030001}{{\em
  Phys. Rev. D} {\bfseries 98} (Aug, 2018) 030001}.

\bibitem{vanDerMeer}
S.~Van Der~Meer, ``{Calibration of the effective beam height in the ISR}'',
  {\em ISR-PO/68-31, KEK68-64} (1968) .

\bibitem{ALICE-PUBLIC-2018-014}
{\bfseries ALICE} Collaboration, ``{ALICE 2017 luminosity determination for pp
  collisions at $\sqrt{s}$ = 5 TeV}'', {\em ALICE-PUBLIC-2018-014} (Nov, 2018)
  . \url{http://cds.cern.ch/record/2648933}.

\bibitem{ALICE-PUBLIC-2017-002}
{\bfseries ALICE} Collaboration, ``{ALICE luminosity determination for pp
  collisions at $\sqrt{s}=8$ TeV}'', {\em ALICE-PUBLIC-2017-002} (Mar, 2017) .
  \url{https://cds.cern.ch/record/2255216}.

\bibitem{ALICE-PUBLIC-2016-002}
{\bfseries ALICE} Collaboration, ``{ALICE luminosity determination for pp
  collisions at $\sqrt{s}=13$ TeV}'', {\em ALICE-PUBLIC-2016-002} (Jun, 2016) .
  \url{https://cds.cern.ch/record/2160174}.

\bibitem{Field:2008zz}
R.~Field, ``{Physics at the Tevatron}'',
{\em Acta Phys. Polon.} {\bfseries B39} (2008) 2611--2672.

\bibitem{2004AcPPB.35.433B}
C.~M. {Buttar}, D.~{Clements}, I.~{Dawson}, and A.~{Moraes}, ``{Simulations of
  minimum bias events and the underlying event, MC tuning and predictions for
  the LHC}'',
{\em Acta Phys. Pol. B} {\bfseries 35} (2004) 433.

\bibitem{Skands:2010ak}
P.~Z. Skands, ``{Tuning Monte Carlo Generators: The Perugia Tunes}'',
  \href{http://dx.doi.org/10.1103/PhysRevD.82.074018}{{\em Phys. Rev.}
  {\bfseries D82} (2010) 074018},
\href{http://arxiv.org/abs/1005.3457}{{\ttfamily arXiv:1005.3457 [hep-ph]}}.

\bibitem{Tsallis:1987eu}
C.~Tsallis, ``{Possible Generalization of Boltzmann-Gibbs Statistics}'',
\href{http://dx.doi.org/10.1007/BF01016429}{{\em J. Statist. Phys.} {\bfseries
  52} (1988) 479--487}.

\bibitem{Abelev:2006cs}
{\bfseries STAR} Collaboration, B.~I. Abelev {\em et~al.}, ``{Strange particle
  production in p+p collisions at $\sqrt{s}$ = 200 GeV}'',
  \href{http://dx.doi.org/10.1103/PhysRevC.75.064901}{{\em Phys. Rev.}
  {\bfseries C75} (2007) 064901},
\href{http://arxiv.org/abs/nucl-ex/0607033}{{\ttfamily arXiv:nucl-ex/0607033
  [nucl-ex]}}.

\bibitem{Acharya:2019qge}
{\bfseries ALICE} Collaboration, S.~Acharya {\em et~al.}, ``{Evidence of
  rescattering effect in Pb-Pb collisions at the LHC through production of
  $\rm{K}^{*}(892)^{0}$ and $\phi(1020)$ mesons}'',
  \href{http://dx.doi.org/10.1016/j.physletb.2020.135225}{{\em Phys. Lett.}
  {\bfseries B802} (2020) 135225},
\href{http://arxiv.org/abs/1910.14419}{{\ttfamily arXiv:1910.14419 [nucl-ex]}}.

\bibitem{Acharya:2019wyb}
{\bfseries ALICE} Collaboration, S.~Acharya {\em et~al.},
  ``{$\rm{K}^{*}(\rm{892})^{0}$ and $\phi(1020)$ production at midrapidity in
  pp collisions at $\sqrt{s}$ = 8 TeV}'',
  \href{http://dx.doi.org/10.1103/PhysRevC.102.024912}{{\em Phys. Rev. C}
  {\bfseries 102} no.~2, (2020) 024912},
  \href{http://arxiv.org/abs/1910.14410}{{\ttfamily arXiv:1910.14410
  [nucl-ex]}}.

\bibitem{Acharya:2019bli}
{\bfseries ALICE} Collaboration, S.~Acharya {\em et~al.}, ``{Multiplicity
  dependence of K*(892)$^{0}$ and $\phi$(1020) production in pp collisions at
  $\sqrt {s}$ =13 TeV}'',
  \href{http://dx.doi.org/10.1016/j.physletb.2020.135501}{{\em Phys. Lett. B}
  {\bfseries 807} (2020) 135501},
  \href{http://arxiv.org/abs/1910.14397}{{\ttfamily arXiv:1910.14397
  [nucl-ex]}}.

\bibitem{Loizides:2017ack}
C.~Loizides, J.~Kamin, and D.~d'Enterria, ``{Improved Monte Carlo Glauber
  predictions at present and future nuclear colliders}'',
  \href{http://dx.doi.org/10.1103/PhysRevC.97.054910,
  10.1103/PhysRevC.99.019901}{{\em Phys. Rev.} {\bfseries C97} no.~5, (2018)
  054910}, \href{http://arxiv.org/abs/1710.07098}{{\ttfamily arXiv:1710.07098
  [nucl-ex]}}.
[erratum: Phys. Rev.C99,no.1,019901(2019)].

\end{thebibliography}\endgroup

\newpage
\appendix
\section{The ALICE Collaboration}
\label{app:collab}
%
\begingroup
\small
\begin{flushleft}
S.~Acharya$^{\rm 142}$, 
D.~Adamov\'{a}$^{\rm 97}$, 
A.~Adler$^{\rm 75}$, 
J.~Adolfsson$^{\rm 82}$, 
G.~Aglieri Rinella$^{\rm 35}$, 
M.~Agnello$^{\rm 31}$, 
N.~Agrawal$^{\rm 55}$, 
Z.~Ahammed$^{\rm 142}$, 
S.~Ahmad$^{\rm 16}$, 
S.U.~Ahn$^{\rm 77}$
% ...

\section*{Affiliation notes}

$^{\rm I}$ Deceased\\
$^{\rm II}$ Also at: Italian National Agency for New Technologies, Energy and Sustainable Economic Development (ENEA), Bologna, Italy\\
$^{\rm III}$ Also at: Dipartimento DET del Politecnico di Torino, Turin, Italy\\
$^{\rm IV}$ Also at: M.V. Lomonosov Moscow State University, D.V. Skobeltsyn Institute of Nuclear, Physics, Moscow, Russia\\
$^{\rm V}$ Also at: Institute of Theoretical Physics, University of Wroclaw, Poland\\

\section*{Collaboration Institutes}

$^{1}$ A.I. Alikhanyan National Science Laboratory (Yerevan Physics Institute) Foundation, Yerevan, Armenia\\
$^{2}$ AGH University of Science and Technology, Cracow, Poland\\
$^{3}$ Bogolyubov Institute for Theoretical Physics, National Academy of Sciences of Ukraine, Kiev, Ukraine\\
$^{4}$ Bose Institute, Department of Physics  and Centre for Astroparticle Physics and Space Science (CAPSS), Kolkata, India\\
$^{5}$ Budker Institute for Nuclear Physics, Novosibirsk, Russia\\
$^{6}$ California Polytechnic State University, San Luis Obispo, California, United States\\
$^{7}$ Central China Normal University, Wuhan, China\\
$^{8}$ Centro de Aplicaciones Tecnol\'{o}gicas y Desarrollo Nuclear (CEADEN), Havana, Cuba\\
$^{9}$ Centro de Investigaci\'{o}n y de Estudios Avanzados (CINVESTAV), Mexico City and M\'{e}rida, Mexico\\
$^{10}$ Chicago State University, Chicago, Illinois, United States\\
$^{11}$ China Institute of Atomic Energy, Beijing, China\\
$^{12}$ Chungbuk National University, Cheongju, Republic of Korea\\
$^{13}$ Comenius University Bratislava, Faculty of Mathematics, Physics and Informatics, Bratislava, Slovakia\\
$^{14}$ COMSATS University Islamabad, Islamabad, Pakistan\\
$^{15}$ Creighton University, Omaha, Nebraska, United States\\
$^{16}$ Department of Physics, Aligarh Muslim University, Aligarh, India\\
$^{17}$ Department of Physics, Pusan National University, Pusan, Republic of Korea\\
$^{18}$ Department of Physics, Sejong University, Seoul, Republic of Korea\\
$^{19}$ Department of Physics, University of California, Berkeley, California, United States\\
$^{20}$ Department of Physics, University of Oslo, Oslo, Norway\\
$^{21}$ Department of Physics and Technology, University of Bergen, Bergen, Norway\\
$^{22}$ Dipartimento di Fisica dell'Universit\`{a} 'La Sapienza' and Sezione INFN, Rome, Italy\\
$^{23}$ Dipartimento di Fisica dell'Universit\`{a} and Sezione INFN, Cagliari, Italy\\
$^{24}$ Dipartimento di Fisica dell'Universit\`{a} and Sezione INFN, Trieste, Italy\\
$^{25}$ Dipartimento di Fisica dell'Universit\`{a} and Sezione INFN, Turin, Italy\\
$^{26}$ Dipartimento di Fisica e Astronomia dell'Universit\`{a} and Sezione INFN, Bologna, Italy\\
$^{27}$ Dipartimento di Fisica e Astronomia dell'Universit\`{a} and Sezione INFN, Catania, Italy\\
$^{28}$ Dipartimento di Fisica e Astronomia dell'Universit\`{a} and Sezione INFN, Padova, Italy\\
$^{29}$ Dipartimento di Fisica e Nucleare e Teorica, Universit\`{a} di Pavia, Pavia, Italy\\
$^{30}$ Dipartimento di Fisica `E.R.~Caianiello' dell'Universit\`{a} and Gruppo Collegato INFN, Salerno, Italy\\
$^{31}$ Dipartimento DISAT del Politecnico and Sezione INFN, Turin, Italy\\
$^{32}$ Dipartimento di Scienze e Innovazione Tecnologica dell'Universit\`{a} del Piemonte Orientale and INFN Sezione di Torino, Alessandria, Italy\\
$^{33}$ Dipartimento di Scienze MIFT, Universit\`{a} di Messina, Messina, Italy\\
$^{34}$ Dipartimento Interateneo di Fisica `M.~Merlin' and Sezione INFN, Bari, Italy\\
$^{35}$ European Organization for Nuclear Research (CERN), Geneva, Switzerland\\
$^{36}$ Faculty of Electrical Engineering, Mechanical Engineering and Naval Architecture, University of Split, Split, Croatia\\
$^{37}$ Faculty of Engineering and Science, Western Norway University of Applied Sciences, Bergen, Norway\\
$^{38}$ Faculty of Nuclear Sciences and Physical Engineering, Czech Technical University in Prague, Prague, Czech Republic\\
$^{39}$ Faculty of Science, P.J.~\v{S}af\'{a}rik University, Ko\v{s}ice, Slovakia\\
$^{40}$ Frankfurt Institute for Advanced Studies, Johann Wolfgang Goethe-Universit\"{a}t Frankfurt, Frankfurt, Germany\\
$^{41}$ Fudan University, Shanghai, China\\
$^{42}$ Gangneung-Wonju National University, Gangneung, Republic of Korea\\
$^{43}$ Gauhati University, Department of Physics, Guwahati, India\\
$^{44}$ Helmholtz-Institut f\"{u}r Strahlen- und Kernphysik, Rheinische Friedrich-Wilhelms-Universit\"{a}t Bonn, Bonn, Germany\\
$^{45}$ Helsinki Institute of Physics (HIP), Helsinki, Finland\\
$^{46}$ High Energy Physics Group,  Universidad Aut\'{o}noma de Puebla, Puebla, Mexico\\
$^{47}$ Hiroshima University, Hiroshima, Japan\\
$^{48}$ Hochschule Worms, Zentrum  f\"{u}r Technologietransfer und Telekommunikation (ZTT), Worms, Germany\\
$^{49}$ Horia Hulubei National Institute of Physics and Nuclear Engineering, Bucharest, Romania\\
$^{50}$ Indian Institute of Technology Bombay (IIT), Mumbai, India\\
$^{51}$ Indian Institute of Technology Indore, Indore, India\\
$^{52}$ Indonesian Institute of Sciences, Jakarta, Indonesia\\
$^{53}$ INFN, Laboratori Nazionali di Frascati, Frascati, Italy\\
$^{54}$ INFN, Sezione di Bari, Bari, Italy\\
$^{55}$ INFN, Sezione di Bologna, Bologna, Italy\\
$^{56}$ INFN, Sezione di Cagliari, Cagliari, Italy\\
$^{57}$ INFN, Sezione di Catania, Catania, Italy\\
$^{58}$ INFN, Sezione di Padova, Padova, Italy\\
$^{59}$ INFN, Sezione di Roma, Rome, Italy\\
$^{60}$ INFN, Sezione di Torino, Turin, Italy\\
$^{61}$ INFN, Sezione di Trieste, Trieste, Italy\\
$^{62}$ Inha University, Incheon, Republic of Korea\\
$^{63}$ Institute for Gravitational and Subatomic Physics (GRASP), Utrecht University/Nikhef, Utrecht, Netherlands\\
$^{64}$ Institute for Nuclear Research, Academy of Sciences, Moscow, Russia\\
$^{65}$ Institute of Experimental Physics, Slovak Academy of Sciences, Ko\v{s}ice, Slovakia\\
$^{66}$ Institute of Physics, Homi Bhabha National Institute, Bhubaneswar, India\\
$^{67}$ Institute of Physics of the Czech Academy of Sciences, Prague, Czech Republic\\
$^{68}$ Institute of Space Science (ISS), Bucharest, Romania\\
$^{69}$ Institut f\"{u}r Kernphysik, Johann Wolfgang Goethe-Universit\"{a}t Frankfurt, Frankfurt, Germany\\
$^{70}$ Instituto de Ciencias Nucleares, Universidad Nacional Aut\'{o}noma de M\'{e}xico, Mexico City, Mexico\\
$^{71}$ Instituto de F\'{i}sica, Universidade Federal do Rio Grande do Sul (UFRGS), Porto Alegre, Brazil\\
$^{72}$ Instituto de F\'{\i}sica, Universidad Nacional Aut\'{o}noma de M\'{e}xico, Mexico City, Mexico\\
$^{73}$ iThemba LABS, National Research Foundation, Somerset West, South Africa\\
$^{74}$ Jeonbuk National University, Jeonju, Republic of Korea\\
$^{75}$ Johann-Wolfgang-Goethe Universit\"{a}t Frankfurt Institut f\"{u}r Informatik, Fachbereich Informatik und Mathematik, Frankfurt, Germany\\
$^{76}$ Joint Institute for Nuclear Research (JINR), Dubna, Russia\\
$^{77}$ Korea Institute of Science and Technology Information, Daejeon, Republic of Korea\\
$^{78}$ KTO Karatay University, Konya, Turkey\\
$^{79}$ Laboratoire de Physique des 2 Infinis, Ir\`{e}ne Joliot-Curie, Orsay, France\\
$^{80}$ Laboratoire de Physique Subatomique et de Cosmologie, Universit\'{e} Grenoble-Alpes, CNRS-IN2P3, Grenoble, France\\
$^{81}$ Lawrence Berkeley National Laboratory, Berkeley, California, United States\\
$^{82}$ Lund University Department of Physics, Division of Particle Physics, Lund, Sweden\\
$^{83}$ Moscow Institute for Physics and Technology, Moscow, Russia\\
$^{84}$ Nagasaki Institute of Applied Science, Nagasaki, Japan\\
$^{85}$ Nara Women{'}s University (NWU), Nara, Japan\\
$^{86}$ National and Kapodistrian University of Athens, School of Science, Department of Physics , Athens, Greece\\
$^{87}$ National Centre for Nuclear Research, Warsaw, Poland\\
$^{88}$ National Institute of Science Education and Research, Homi Bhabha National Institute, Jatni, India\\
$^{89}$ National Nuclear Research Center, Baku, Azerbaijan\\
$^{90}$ National Research Centre Kurchatov Institute, Moscow, Russia\\
$^{91}$ Niels Bohr Institute, University of Copenhagen, Copenhagen, Denmark\\
$^{92}$ Nikhef, National institute for subatomic physics, Amsterdam, Netherlands\\
$^{93}$ NRC Kurchatov Institute IHEP, Protvino, Russia\\
$^{94}$ NRC \guillemotleft Kurchatov\guillemotright  Institute - ITEP, Moscow, Russia\\
$^{95}$ NRNU Moscow Engineering Physics Institute, Moscow, Russia\\
$^{96}$ Nuclear Physics Group, STFC Daresbury Laboratory, Daresbury, United Kingdom\\
$^{97}$ Nuclear Physics Institute of the Czech Academy of Sciences, \v{R}e\v{z} u Prahy, Czech Republic\\
$^{98}$ Oak Ridge National Laboratory, Oak Ridge, Tennessee, United States\\
$^{99}$ Ohio State University, Columbus, Ohio, United States\\
$^{100}$ Petersburg Nuclear Physics Institute, Gatchina, Russia\\
$^{101}$ Physics department, Faculty of science, University of Zagreb, Zagreb, Croatia\\
$^{102}$ Physics Department, Panjab University, Chandigarh, India\\
$^{103}$ Physics Department, University of Jammu, Jammu, India\\
$^{104}$ Physics Department, University of Rajasthan, Jaipur, India\\
$^{105}$ Physikalisches Institut, Eberhard-Karls-Universit\"{a}t T\"{u}bingen, T\"{u}bingen, Germany\\
$^{106}$ Physikalisches Institut, Ruprecht-Karls-Universit\"{a}t Heidelberg, Heidelberg, Germany\\
$^{107}$ Physik Department, Technische Universit\"{a}t M\"{u}nchen, Munich, Germany\\
$^{108}$ Politecnico di Bari and Sezione INFN, Bari, Italy\\
$^{109}$ Research Division and ExtreMe Matter Institute EMMI, GSI Helmholtzzentrum f\"ur Schwerionenforschung GmbH, Darmstadt, Germany\\
$^{110}$ Russian Federal Nuclear Center (VNIIEF), Sarov, Russia\\
$^{111}$ Saha Institute of Nuclear Physics, Homi Bhabha National Institute, Kolkata, India\\
$^{112}$ School of Physics and Astronomy, University of Birmingham, Birmingham, United Kingdom\\
$^{113}$ Secci\'{o}n F\'{\i}sica, Departamento de Ciencias, Pontificia Universidad Cat\'{o}lica del Per\'{u}, Lima, Peru\\
$^{114}$ St. Petersburg State University, St. Petersburg, Russia\\
$^{115}$ Stefan Meyer Institut f\"{u}r Subatomare Physik (SMI), Vienna, Austria\\
$^{116}$ SUBATECH, IMT Atlantique, Universit\'{e} de Nantes, CNRS-IN2P3, Nantes, France\\
$^{117}$ Suranaree University of Technology, Nakhon Ratchasima, Thailand\\
$^{118}$ Technical University of Ko\v{s}ice, Ko\v{s}ice, Slovakia\\
$^{119}$ The Henryk Niewodniczanski Institute of Nuclear Physics, Polish Academy of Sciences, Cracow, Poland\\
$^{120}$ The University of Texas at Austin, Austin, Texas, United States\\
$^{121}$ Universidad Aut\'{o}noma de Sinaloa, Culiac\'{a}n, Mexico\\
$^{122}$ Universidade de S\~{a}o Paulo (USP), S\~{a}o Paulo, Brazil\\
$^{123}$ Universidade Estadual de Campinas (UNICAMP), Campinas, Brazil\\
$^{124}$ Universidade Federal do ABC, Santo Andre, Brazil\\
$^{125}$ University of Cape Town, Cape Town, South Africa\\
$^{126}$ University of Houston, Houston, Texas, United States\\
$^{127}$ University of Jyv\"{a}skyl\"{a}, Jyv\"{a}skyl\"{a}, Finland\\
$^{128}$ University of Kansas, Lawrence, Kansas, United States\\
$^{129}$ University of Liverpool, Liverpool, United Kingdom\\
$^{130}$ University of Science and Technology of China, Hefei, China\\
$^{131}$ University of South-Eastern Norway, Tonsberg, Norway\\
$^{132}$ University of Tennessee, Knoxville, Tennessee, United States\\
$^{133}$ University of the Witwatersrand, Johannesburg, South Africa\\
$^{134}$ University of Tokyo, Tokyo, Japan\\
$^{135}$ University of Tsukuba, Tsukuba, Japan\\
$^{136}$ Universit\'{e} Clermont Auvergne, CNRS/IN2P3, LPC, Clermont-Ferrand, France\\
$^{137}$ Universit\'{e} de Lyon, CNRS/IN2P3, Institut de Physique des 2 Infinis de Lyon , Lyon, France\\
$^{138}$ Universit\'{e} de Strasbourg, CNRS, IPHC UMR 7178, F-67000 Strasbourg, France, Strasbourg, France\\
$^{139}$ Universit\'{e} Paris-Saclay Centre d'Etudes de Saclay (CEA), IRFU, D\'{e}partment de Physique Nucl\'{e}aire (DPhN), Saclay, France\\
$^{140}$ Universit\`{a} degli Studi di Foggia, Foggia, Italy\\
$^{141}$ Universit\`{a} di Brescia, Brescia, Italy\\
$^{142}$ Variable Energy Cyclotron Centre, Homi Bhabha National Institute, Kolkata, India\\
$^{143}$ Warsaw University of Technology, Warsaw, Poland\\
$^{144}$ Wayne State University, Detroit, Michigan, United States\\
$^{145}$ Westf\"{a}lische Wilhelms-Universit\"{a}t M\"{u}nster, Institut f\"{u}r Kernphysik, M\"{u}nster, Germany\\
$^{146}$ Wigner Research Centre for Physics, Budapest, Hungary\\
$^{147}$ Yale University, New Haven, Connecticut, United States\\
$^{148}$ Yonsei University, Seoul, Republic of Korea\\

\bigskip 

\end{flushleft} 
\endgroup  
\end{document}